\documentclass[12pt, a4paper]{article}

\pdfoutput=1

\usepackage{amsmath}
\usepackage{amsfonts}
\usepackage{amssymb}
\usepackage{graphicx, rotating}
\usepackage{epstopdf}
\usepackage{epsfig}
\usepackage{latexsym}
\usepackage{subfigure}
\usepackage{graphicx}
\usepackage{color}
\usepackage{amsmath,amssymb}
\usepackage{cite}
\usepackage{slashed}
\usepackage{hyperref}
\usepackage[export]{adjustbox}

\hypersetup{colorlinks, citecolor=bluscuro, linkcolor=black, urlcolor=bluscuro}
\definecolor{rossos}{cmyk}{0,1,1,0.55}
\definecolor{bluscuro}{rgb}{0.15, 0.2, .85}
\definecolor{bluchiaro}{cmyk}{1,.3,0.,0.1}

\newcommand{\eq}[1]{Eq.~(\ref{#1})}

\newcommand{\Lag}{\mathcal{L}}

\newcommand{\nn}{\nonumber}

\newcommand{\be}{\begin{equation}}
\newcommand{\ee}{\end{equation}}
\newcommand{\bea}{\begin{eqnarray}}
\newcommand{\eea}{\end{eqnarray}}
\newcommand{\bc}{\begin{center}}
\newcommand{\ec}{\end{center}}
\newcommand{\Eq}[1]{Eq.\!~(\ref{#1})}
\newcommand{\Fig}[1]{Fig.\!~\ref{#1}}
\newcommand{\Eqs}[1]{Eqs.\!~(\ref{#1})}

\newcommand{\blu}{\color{blue}}

\newcommand{\N}{\mathcal{N}} 

\newcommand{\TeV}{\,\mathrm{TeV}}
\newcommand{\GeV}{\,\mathrm{GeV}}

\newcommand{\beq}{\begin{equation}}
\newcommand{\eeq}{\end{equation}}
\newcommand{\bag}{\begin{align}}
\newcommand{\eag}{\end{align}}

\def\lra#1{\overset{\text{\scriptsize$\leftrightarrow$}}{#1}}
\newcommand{\Op}{{\mathcal O}}

\newcommand{\drawsquare}[2]{\hbox{%
\rule{#2pt}{#1pt}\hskip-#2pt
\rule{#1pt}{#2pt}\hskip-#1pt
\rule[#1pt]{#1pt}{#2pt}}\rule[#1pt]{#2pt}{#2pt}\hskip-#2pt
\rule{#2pt}{#1pt}}

\newcommand{\Yfund}{\drawsquare{7}{0.6}}

\begin{document}

\vspace*{-2cm}
\begin{flushright}
{\small Saclay-t17/081}\\
{\small CERN-TH-2017-119}\\
{\small SISSA 30/2017/FISI}
\end{flushright}

\begin{center}
\vspace*{15mm}

\vspace{1cm}
{\Large \bf 
The Other Fermion Compositeness
}
\vspace{1.4cm}

{Brando Bellazzini$\,^{a,\, b}$, Francesco~Riva$\,^{c}$, Javi Serra$\,^{c}$ and Francesco Sgarlata$\,^{d}$}

 \vspace*{.5cm} 
\begin{footnotesize}
\begin{it}
 $^a$ Institut de Physique Th\'eorique, Universit\'e Paris Saclay, CEA, CNRS, F-91191 Gif-sur-Yvette, France\\
$^b$ Dipartimento di Fisica e Astronomia, Universit\'a di Padova,  Via Marzolo 8, I-35131 Padova, Italy\\
$^c$ CERN, Theory Department, CH-1211 Geneve 23, Switzerland \\
$^d$  SISSA International School for Advanced Studies, Via Bonomea 265, 34136, Trieste, Italy
\end{it}
\end{footnotesize}

\vspace*{.2cm} 

\vspace*{10mm} 
\begin{abstract}\noindent\normalsize
We discuss the only two viable realizations of fermion compositeness 
{described by a calculable relativistic effective field theory} consistent with unitarity, crossing symmetry and analyticity: \textit{chiral-compositeness} vs \textit{goldstino-compositeness}. We construct the effective theory of $\mathcal{N}$ Goldstini and show how the Standard Model can emerge from this dynamics. We present new bounds on either type of compositeness, for quarks and leptons, using dilepton searches at LEP, dijets at the LHC, as well as low-energy observables and precision measurements. Remarkably, a scale of compositeness for Goldstino-like electrons in the 2 TeV range is compatible with present data, and so are Goldstino-like first generation quarks with a compositeness scale in the 10 TeV range. Moreover, assuming maximal $R$-symmetry, goldstino-compositeness of both right- and left-handed quarks predicts exotic spin-1/2 colored sextet particles that are potentially within the reach of the LHC.
\end{abstract}

\end{center}

\newpage

\section{Motivation}

The LHC defines the frontiers of our exploration of the universe at microscopic scales.
Its primary focus so far has been the search for new physics in the form of narrow resonances, often associated with weakly coupled (calculable) physics beyond the Standard Model (SM).
The incredible amount of data that is and will be accumulated at the LHC gives access to unprecedented accuracy in our knowledge of SM processes. A crucial question is then to understand what kind of physics can be tested with all this new information.
Amusingly, strongly (as opposed to weakly) coupled dynamics, associated with the compositeness of SM particles, can produce sizable deviations in their interactions at high energy and become the principal target of these SM precision tests \cite{Giudice:2007fh,Contino:2010mh,Domenech:2012ai,deBlas:2013qqa,Dror:2015nkp,Liu:2016idz,Contino:2016jqw,Falkowski:2016cxu,Farina:2016rws,Biekoetter:2014jwa}.

In this article we present a perspective on how the SM can emerge as the result of such a strong dynamics, a perspective on how high-energy compositeness can appear  to a low-energy observer.
We are interested in situations where the Compton wavelength of a particle $\sim m^{-1}$ is much larger than the microscopic scale $\sim m_*^{-1}$ of the confining theory, so that it effectively appears elementary at energies of order its mass.%
\footnote{Note therefore that we are not concerned here with compositeness of the type the proton exhibits for instance. This would give rise to large non-standard effects for the SM fermions already at energies comparable to their mass, effects which have been long ruled out experimentally. Instead our focus is on the class of compositeness that can only be efficiently probed at higher energies, in the spirit of Ref.~\cite{Eichten:1983hw}.}
A hierarchy between these different scales can be naturally generated by the presence of certain approximate symmetries that characterize the strong dynamics but are explicitly broken by $m$ and by the other SM interactions. Notable examples are non-linearly realized symmetries that protect the mass of Nambu-Goldstone modes, or chiral symmetry that protects that of fermions.
In practice, the imprint of compositeness can be captured in an effective Lagrangian by irrelevant but strong operators of dimension $D>4$, weakly deformed by marginal or relevant (SM-like) interactions with $D \leq 4$. At low energy only the latter survive and particles look SM-like, while at high-energy the former grow and the new strong interaction is gradually revealed.

The more irrelevant the new interactions, the more this scenario  would resemble the SM at low energy, simply because the effect of compositeness would vanish more rapidly as the energy that is used to probe it decreases. It is therefore crucial to understand how well can a framework of compositeness in the multi-TeV region (accessible at present or near-future colliders) fake the SM: how irrelevant can these new interactions be? 
From a low-energy perspective one would see no obstruction in arbitrarily \emph{soft} interactions, but dispersion relations can be used to show that the basic requirement of unitarity of the microscopic (UV) theory, imposes certain positivity constraints that guarantee the existence of unsuppressed dimension-8 operators \cite{Bellazzini:2016xrt,Bellazzini:2014waa,Adams:2006sv}. Given the SM field content and limiting our discussion to lepton and baryon number preserving effects, this implies that the \textit{leading} BSM effects must be characterized either by operators of dimension-6 or by dimension-8, but not higher. 
While most of the literature focuses on the former case, in this article we discuss instead the latter situation in which the leading BSM interactions saturate the unitarity bound and arise at dimension-8: these are theories that flow maximally fast to the SM as energy is decreased. 

In fact, systems with very soft behavior have been studied in the context of scalars including the Higgs \cite{Liu:2016idz,Bellazzini:2015cgj}, and vectors \cite{Liu:2016idz}.
 In this article we focus on fermions, and go beyond the traditional paradigm where fermion compositeness relies on chiral symmetry and the associated four-fermion dimension-6 operators. 
 We consider fermions whose composite nature is dominantly captured by dimension-8 operators (four-fermion with two extra derivatives) via a non-linearly realized supersymmetry that forbids operators with $D<8$. That is, (some of) the SM fermions are pseudo-Goldstini, in a modern generalization of the Akulov-Volkov theory for the neutrino \cite{Volkov:1973ix,Bardeen:1981df}. For this reason we dub this power-counting  \textit{goldstino-compositeness}, as opposed to the well-known \textit{chiral-compositeness} based on chiral symmetry. 

{Present and future colliders have incredible sensitivity to modifications of the properties of light fermions.
We find this a strong motivation to understand fermion compositeness from both a theoretical and phenomenological point of view and to assess in which ways processes involving fermions can carry information about any structurally robust BSM dynamics,  even if exotic.
This ambitious task is simplified by our observation that there exist only two patterns of fermion compositeness compatible with prime principles. }
The traditional chiral-compositeness of fermions, which gives rise to dimension-6  operators, is incredibly well constrained by LHC data for the light quarks, leaving no hope of ever testing it directly on an earth-based collider (see section~\ref{sec:coll} or Ref.~\cite{Domenech:2012ai}). 
On the other hand, non-linearly realized supersymmetry, i.e.~goldstino-compositeness, provides an alternative explanation for the small fermion masses, while allowing large effects in the more irrelevant dimension-8 operators, whose collider constraints, as we will show here, are much less severe. 
Interestingly, the fast decoupling of the Goldstino interactions becomes distinctive as well when considering diboson production, whose leading modifications come from dimension-8 operators when the Higgs or the transverse gauge bosons are also composite.

This paper is organised as follows. In section~\ref{sec:UV} we present our picture for goldstino-compositeness based on approximate extended supersymmetries. 
We construct the effective theory of Goldstini in section~\ref{sec:EFT}. In this theoretically oriented section we discuss the geometry of the coset space associated to supersymmetry breaking, the general interactions of the Goldsitini, and the embedding of quarks and leptons.
In section~\ref{sec:break} we understand the deformations induced by the explicit breaking of supersymmetry. 
The phenomenology of our scenario is covered in sections~\ref{sec:coll} and \ref{sec:exoticXY}, where we discuss, respectively, the $2 \to 2$ scattering process that best probe the compositeness of quarks and leptons and the exotic ``quixes'' that are predicted by maximal $R$-symmetry.

\section{Pseudo-Goldstini}
\label{sec:UV}

We consider a strongly interacting supersymmetric sector. It confines breaking SUSY spontaneously and leaving only massless Goldstini in the infrared (IR) spectrum, that we identify with (some of) the SM fermions; in addition to kinetic terms, Goldstini have self-interactions that start at dimension-8. This picture is deformed by non-supersymmetric SM couplings, which we take to be small compared with the new sector's coupling, thus we treat them as  perturbative deformations of the exact Goldstino limit.%
\footnote{Another interesting application of this ideology is Goldstino Dark Matter \cite{Bruggisser:2016ixa}.} 
Yet, these small effects are marginal and at small enough energy they dominate, while at high energy Goldstino self-interactions  become more important.

We will study scenarios where either one, several, or even all of the quarks and leptons are (pseudo-)Goldstini, implying the existence of $1\leq {\cal N}\leq 84$ supersymmetries (see section~\ref{sec:emb}) -- numbers at which the reader might be willing to raise an eyebrow. It is indeed  well known that complete massless supermultiplets (i.e.~within a linearly realized SUSY with massless particles) and ${\cal N}> 8$ supercharges imply the existence of \emph{massless} higher-spin states which are pathological, in flat space, on very general grounds \cite{Weinberg:1964ew,Coleman:1967ad,Porrati:2008rm} (see e.g.~\cite{Bekaert:2010hw} for a review). Yet, the roots of these arguments are based on the IR properties of the states, i.e.~the soft scattering limits of the $S$-matrix elements or the existence of global charges, which is exactly the regime where one is sensitive to the spontaneous symmetry breaking effects, i.e.~the soft masses.
A spontaneously broken extended SUSY does not predict those massless higher spins at all.%
\footnote{Incidentally, in our framework neither the gauge bosons nor the graviton are actually part of the strong sector (more on this below), evading even more the existence of massless supermultiplets.} 
In non-linearly realized SUSY, the would-be higher-spin superpartners are actually multi-particle states obtained by including Goldstino insertions that do in fact raise/lower the spin, but without producing single particle states.%
\footnote{This is made manifest for instance in the constrained superfield formalism \cite{Casalbuoni:1988xh,Komargodski:2009rz}, where the scalar partner of the Goldstino is a pair of Goldstini, $X=\chi^2+\theta\chi+\theta^2 \mathcal{F}$.} 
 
 One can  take a step further and argue that the very existence of ${\cal N}> 8$ implies  the only consistent phase of SUSY in flat space is the broken one, which in turn requires the existence of fermions much lighter than the cutoff, as we observe in nature. There could then be a fascinating link between the (nearly) vanishing cosmological constant and the existence of (light) spin-1/2 fermions in the spectrum. Moreover, it could be  possible that a UV completion in the unbroken phase needs to be formulated in AdS space, given the need to inject a positive contribution to the vacuum energy to move to the broken phase. Notice that massless higher spins in AdS pose no problem \cite{Vasiliev:1999ba,Vasiliev:1995dn,Bekaert:2010hw}. 

There is in fact another difficulty with ${\cal N}\geq 8$: the impossibility of interactions of renormalizable dimension $D \leq 4$, in a Lagrangian description. The only interactions that can be written compatibly with these extended supersymmetries are highly irrelevant. This suggests that, if a UV completion exists, is not of a weakly coupled kind and does not necessarily rely on a Lagrangian or effective description. It may well be that a UV completion expressed as local Lagrangian is made of a series of higher-dimension operators that become important at energies of order the higher-spin states mass, yet resulting in a well-defined $S$-matrix at any finite energy. This is somewhat analogous to what happens in Vasiliev's theories in curved AdS space \cite{Vasiliev:1999ba,Vasiliev:1995dn}, where the cutoff is  given by the cosmological constant itself, i.e.~the deepest IR observable, such that theory is never weakly coupled in the UV, where the curvature could in principle be neglected. Another example is the high-energy limit of string theory \cite{Gross:1988ue}, where infinitely many higher-spin conserved currents appear to be restored, corresponding to the tensionless limit.

Alternatively, extended SUSY could just be realized as an emergent symmetry which (re)appears in the IR from UV dynamics that do not necessarily exhibits such a symmetry \cite{Kaplan:1983sk,Strassler:2003ht,Sundrum:2009gv}. Yet another option is to consider several almost sequestered $\N\leq 8$ supersymmetric sectors which are linked by an approximate permutation symmetry, e.g.~$Z_M \rtimes [\N\mbox{-SUSY}]^M$ (where $M$ is a flavor index).\\

In any case, since there are no manifest obstructions to the presence of several supercharges, we assume that a strongly coupled sector that breaks an extended SUSY indeed exists. We call $g_*\lesssim 4\pi$ its strong coupling at the scale of confinement $m_*$. We will consider first the limit of rigid supersymmetry and no gravity $(M_{Pl}\to \infty)$, as well as vanishing SM couplings $g_{SM} \equiv \{g,g^\prime,g_s,Y_\psi\}\to 0$. In this limit we assume SUSY is spontaneously broken with order parameter $F\sim m_*^2/g_*$, resulting in ${\cal N}$ massless Goldstini.

A finite $M_{Pl}$ corresponds to the gauging of Poincar\'e symmetry, which in the spontaneously broken supersymmetric context implies supergravity with massive Gravitini \cite{Volkov:1973jd}. This is not how nature looks like: the SM fermions have spin $1/2$, excluding ${\cal N}$ supergravities.  Instead, finite $M_{Pl}$ with \emph{rigid} supersymmetry introduces \emph{explicit} SUSY breaking (the graviton has no superpartner) that can potentially propagate to the matter sector, giving large contributions to relevant operators (e.g.~$\sim M_{Pl}^2$ contributions to scalar masses) and masses to the Goldstini. However, in theories without elementary scalars \cite{Kaplan:1983sk}, or strongly coupled theories where the $|\phi|^2$ operators have $D\gtrsim 4$ \cite{Strassler:2003ht,Sundrum:2009gv}, SUSY is still approximately preserved. Moreover, an exact ${\cal N}=1$ supergravity in addition to ${\cal N}$ rigid supersymmetries would be enough to protect scalar masses and preserve the necessary amount of SUSY~\cite{ArkaniHamed:2000ds}.

The strongly interacting supersymmetric sector is also coupled to an elementary sector which includes, possibly, but not necessarily, elementary (non-supersymmetric) scalars, gauge bosons and some fermions. The elementary sector then breaks SUSY explicitly, but in a way that can be kept under control, given that $g_{SM}\ll g_*$ is a small perturbation of the undeformed theory, in the spirit of Ref.~\cite{Gherghetta:2003he}. In particular, the SM gauge vectors couple to (part of) the conserved currents in the supersymmetric sector, which are associated to an unbroken $R$-symmetry under which the Goldstini-like SM fermions transform.%
\footnote{There is a priori a different starting point to this construction in which $R$-symmetry is gauged (necessarily together with supergravity) \cite{Freedman:1976uk}. However, in such theories $R$-symmetry is broken at very high scales by the VEV of the superpotential, in order to cancel the cosmological constant; see however \cite{Claudson:1983cr} for a possible caveat.}
We further assume that $R$-symmetry includes (part of) the SM flavor group, broken only by the Yukawa couplings to the Higgs boson. This construction reproduces a form of MFV~\cite{DAmbrosio:2002vsn} that successfully surpasses flavor constraints even for a low SUSY-breaking scale.

\section{The Effective Goldstini Theory}
\label{sec:EFT}

Following the outline in the previous section, we first construct the Goldstini EFT, in the limit where (rigid) explicit SUSY breaking effects vanish, that is $g_{SM}\to 0$ and $M_{Pl}\to\infty$. 
{This section is rather formal: the results relevant for the rest of the paper are summarized in Tables \ref{tab1:indepcoup}, \ref{tab2:modedepcoup} and \ref{tab:embeddings}.}

\subsection{The geometry of $\N$ Goldstini}

As for every Goldstone field, the Goldstini $\chi_i$ can be chosen to parametrize the coset space associated to spontaneous SUSY breaking \cite{Coleman:1969sm,Callan:1969sn,Volkov:1973vd,Ivanov:1975zq,Volkov:1973ix}.%
\footnote{Different and yet equivalent parametrizations are often adopted for the Goldstini, like the one provided by the constrained superfield formalism, see e.g.~\cite{Komargodski:2009rz,Cribiori:2016hdz} and references therein. Constrained superfields are convenient whenever the Goldstini couple to IR fields that fill complete SUSY multiplets, i.e. in linear SUSY representations. This is never the case in our EFT.}
They provide a map from the spacetime to an element of the SUSY transformations, $x\rightarrow g(x)$, up to the identification $g(x)\sim g(x)h(x)$ where $h(x)$ is an element of the unbroken symmetry group that includes (homogeneous) Lorentz transformations.%
\footnote{Spacetime translations are effectively considered as broken generators since they are non-linearly realized on spacetime, which is nothing but the coset Poincar\'e/Lorentz, $x\rightarrow x+c$ (see e.g. \cite{Delacretaz:2014oxa}).}
At the Lagrangian level, this identification is realized as usual by constructing a gauge theory invariant under local  Lorentz transformations, a gravity theory of a sort, although with a non-dynamical \textit{composite vielbein} $E_\mu^{\phantom{\mu}a}(\chi)$ made of Goldstini. To this end, we choose the following coset representative
\begin{equation}
\label{defU}
U(x,\chi(x))\equiv e^{i\left(\chi(x) Q+\chi^\dagger(x) Q^\dagger\right)}e^{ix^\mu P_\mu} \, , 
\end{equation}
where $\chi Q$ and $\chi^\dagger Q^\dagger$ are shorthand for $\chi^\alpha_i Q_\alpha^i$ and $\chi_{\dot\alpha}^{\dagger i} Q^{\dagger\,\dot\alpha}_i$ respectively, where the sum over repeated indexes is always understood unless stated otherwise.  In the following, both spinor and flavor contractions are  understood.
The relevant part of the extended SUSY algebra  we consider is 
\begin{equation}
\label{Alg}
\{Q^i_\alpha, Q^{\dagger}_{\dot{\beta}\,j}\} = 2\sigma^\mu_{\alpha\dot\beta}P_\mu \delta^i_j  \, , \qquad
[P_\mu, Q_\alpha^i] = [P_\mu, Q^{\dagger\,i}_{\dot\alpha}]=\{Q^i_\alpha, Q^j_\beta\} = \{Q^{\dagger\,i}_{\dot\alpha}, Q^{\dagger\,j}_{\dot\beta}\}=0 \, , 
\end{equation}
where the latin indexes $i,j=1,\ldots, \N$ label the supercharges, while the greek indexes are spinorial in the $(1/2,0)$ (undotted) or $(0,1/2)$ (dotted) representation of the Lorentz group $SO(1,3)\sim SU(2)\times SU(2)$.  We assume no central charge since, as explained above, we are eventually interested in identifying the unbroken $R$-symmetry $G_R\subseteq U(\N)\sim SU(\N)_R \times U(1)_R$, or some of its subgroups, with the SM flavor and gauge groups. 

Under a global transformation $g$,  the representative element $U$ is moved into another group element of the form $g U=U^\prime h$ which defines the transformation
\begin{equation}
U(x,\chi(x))\rightarrow U^\prime(x^\prime,\chi^\prime(x^\prime))= g U(x,\chi(x)) h^{-1} \, , 
\end{equation}
 for the coset representative, hence for the Goldstini. 
 
The Goldstini transform linearly under the unbroken symmetries. In particular, they transform under Lorentz  as ordinary two-component spinors.%
\footnote{Explicitly,  we have $L U(x,\chi(x))=L e^{i\left(\chi(x) Q+\chi^\dagger(x) Q^\dagger\right)} L^{-1} \left(Le^{ix^\mu P_\mu}L^{-1}\right)L=U^\prime(x^\prime,\chi^\prime(x)) L$ where one can use $L P_\mu L^{-1}= P_\nu \Lambda^{\nu}_{\phantom{\nu}\mu}$, $L Q_\alpha^i L^{-1}= \widetilde{\Lambda}_\alpha^{\phantom{\alpha}\beta} Q_{\beta}^i$, and  $L Q^{\dagger\,\dot\alpha}_i L^{-1}= \widehat{\Lambda}^{\dot\alpha}_{\phantom{\alpha}\dot{\beta}} Q^{\dot\beta}_i$ to get the Lorentz action on the Goldstini,  namely $\chi^{\prime\,\alpha}(x^\prime) =\chi^\beta (x(x^\prime)) \,\widetilde{ \Lambda}_{\beta}^{\phantom{\beta}\alpha}$ with $\widetilde{ \Lambda}_{\beta}^{\phantom{\beta}\alpha}$ the spinorial $(1/2,0)$ representation of $L$.} 
Similarly, for an unbroken $R$-symmetry  $G_R$ the Goldstini $\chi$ and $\chi^\dagger$ carry a linear representation, e.g.~the fundamental representation of maximal $R$-symmetry 
\begin{align}
&[R^a_{SU(\N)_R}, Q^i]=\left(T^a\right)^{i}_{\phantom{i}j} Q^j\,,\qquad  [R_{U(1)_R}, Q^i] = Q^i\,, &\nn\\
&[R^a_{SU(\N)_R}, Q^\dagger_i]=-\left(\overline{T}^a\right)^{i}_{\phantom{i}j} Q^\dagger_j\,,\qquad   [R_{U(1)_R}, Q^\dagger_i] = -Q^\dagger_i \,, &\label{commrel}
\end{align}
where $T^a$ are the $\N^2-1$ traceless and hermitian $\N$--by--$\N$ generators of $SU(\N)_R$. 
 
Under spacetime translations $T=e^{i a^\mu P_\mu}$, $U^\prime=T U$ where $x^\prime=x+a$ and $\chi^\prime(x^\prime)=\chi(x(x^\prime))=\chi(x^\prime-a)$ as usual. 
 
Under the action of the broken SUSY generators $ g_{\xi}=\exp[i\xi_j Q^j+i\xi^{j\dagger} Q^{\dagger}_j]$, we get instead a non-linear realization:  
\begin{align}
\label{defUp}
g_{\xi} \, U(x,\chi(x)) & =U^\prime(x^\prime,\chi^\prime(x^\prime))\,\quad \rightarrow\quad\left\{
\begin{array}{l}\chi^\prime(x^\prime)  =\chi(x(x^\prime))+ \xi \,\\
x^{\prime\mu}(x) =x^{\mu} - i\chi^\dagger(x) \bar{\sigma}^\mu \xi +i \xi^\dagger \bar{\sigma}^\mu \chi(x)\end{array}\right.\,.
\end{align}
Indeed, under an infinitesimal $\xi^i$ it corresponds to the following non-linear action of the SUSY transformations 
\begin{align}
\label{chitpoint}
\chi(x)\rightarrow \chi^\prime(x) & = \chi(x^\prime(x))+\xi=  \chi(x)+\xi -v^\mu(\xi,\chi)\partial_\mu\chi(x)+\ldots
\end{align}
where $v^\mu(\xi,\chi(x))\equiv -i\left(\chi^\dagger(x)\bar{\sigma}^\mu\xi - \xi^\dagger\bar{\sigma}^\mu \chi(x)\right)$.

For matter fields (scalars and spinors)  the non-linear representation is the same as for the Goldstini except that it does not include the Grassmannian shift $\xi$, namely
\begin{equation}
\label{standardrep}
\Phi(x)\rightarrow \Phi^\prime(x)=\Phi(x^\prime(x))=\Phi(x)-v^\mu(\xi,\chi(x))\partial_\mu\Phi(x)+\ldots
\end{equation}

\subsubsection{Covariant derivatives and Maurer-Cartan form}

Notice that derivatives transform non-covariantly, that is differently than the undifferentiated fields.%
\footnote{Explicitly, $\partial_\mu \chi(x)\rightarrow \partial_\mu \chi(x^\prime(x))=(\partial_\nu\chi)(x^\prime(x)) \,\frac{\partial x^{\prime\nu}}{\partial x^\mu} =\partial_\mu\chi(x^\prime(x))-\partial_\nu\chi(x^\prime(x))\, v^\nu(\xi,\partial_\mu\chi(x))+\ldots$ and only the first term on the right-hand side corresponds to the covariant SUSY transformation.} 
In order to deal with proper covariant transformations of derivatives, it is useful to introduce the Maurer-Cartan 1-form in the SUSY algebra 
\begin{equation}
(U^{-1}d U)(x)=idx^\mu E_\mu^{\phantom{\mu}a} \left( P_a+ \nabla_a \chi\, Q+\nabla_a \chi^\dagger\, Q^\dagger\right) \, , 
\end{equation}
where for future convenience we have factored out the coefficient $E_\mu^{\phantom{\mu}a}$ of the momentum 
\begin{align}
\label{Edef}
E_\mu^{\phantom{\mu}a} &= \delta^a_\mu - i\chi^{j\dagger} \bar{\sigma}^a\partial_\mu\chi_j +i\partial_\mu\chi^{j\dagger}\bar{\sigma}^a \chi_j\,,\\
\nabla_a \chi_j & = (E^{-1})_a^{\phantom{a}\mu} \partial_\mu\chi_j \equiv  E_a^{\phantom{a}\mu} \partial_\mu\chi_j \, , 
\end{align}
with $\nabla_a \chi^{j \dagger}=(\nabla_a\chi^j)^\dagger$, and defined the inverse of $E_\mu^{\phantom{\mu}a}$ as 
\begin{equation}
(E^{-1})_a^{\phantom{a}\mu}\equiv  E_a^{\phantom{a}\mu}\,,\qquad E_a^{\phantom{a}\mu} E_\mu^{\phantom{\mu}b}=\delta^b_a\,,\qquad E_\mu^{\phantom{\mu}a}E_a^{\phantom{a}\nu}=\delta^\nu_\mu\,.
\end{equation}
Since the Maurer-Cartan form is invariant, $(U^{-1}d U)^\prime(x^\prime)=(U^{-1}d U)(x)$, its change at a given point is just the contraction with the Jacobian matrix. 
This tells us that the  $E_\mu^{\phantom{\mu}a}$ and $E_a^{\phantom{a}\mu}$
  transform as vielbeins 
\begin{align}
E_\mu^{\phantom{\mu}a}(x)  \rightarrow E_\mu^{\prime\phantom{\mu}a}(x) = \frac{\partial x^{\prime \nu}}{\partial x^\mu}E_\nu^{\phantom{\nu}a}(x^\prime(x)) \,,  \qquad 
E_a^{\phantom{a}\mu}(x)  \rightarrow E_a^{\prime\phantom{a}\mu}(x) = \frac{\partial x^\mu}{\partial x^{\prime \nu}} E_a^{\phantom{a}\nu}(x^\prime(x)) \,.
\end{align}
It is therefore clear that  $\nabla_a\chi$ transforms covariantly, as in Eq.~(\ref{standardrep}),
\begin{equation}
\nabla_a \chi(x)\rightarrow (\nabla_a \chi)^\prime (x)=(\nabla_a \chi)(x^\prime(x))\,.
\end{equation}

Gauge fields behave just as derivatives. The resulting gauge- and SUSY-covariant derivatives read $D_a\equiv\nabla_a- i  \mathbb{A}_a = E_a^{\phantom{a}\mu}\left(\partial_\mu-i A_\mu \right)$.  Similarly, the covariant field strength is 
\begin{equation}
\label{dressingFieldStre}
\mathbb{F}_{ab}= E_a^{\phantom{a}\mu}E_b^{\phantom{b}\nu}\left(\partial_\mu A_\nu -\partial_\nu A_\mu-[A_\mu,A_\nu]\right)\,.
\end{equation} 
In our EFT we gauge an $R$-symmetry, as opposed to an ordinary global symmetry, which forces us to break SUSY explicitly. However the gauge fields themselves could in principle be part of the strong sector, {that is composite} 
\cite{Liu:2016idz}. In this case, one should dress with the vielbeins only the $\partial_{[\mu} A_{\nu]}$ part of the field strength and not its non-abelian part $[A_\mu,A_\nu]$. For elementary gauge fields neither term should be dressed with Goldstini. 

\subsubsection{Effective metric and invariant measure}

From the composite vielbeins one can define a composite effective metric
\begin{equation}
\label{compmetr}
g_{\mu\nu}(x)\equiv E_\mu^{\phantom{\mu}a}(x) E_\nu^{\phantom{\mu}b}(x)\eta_{ab}\rightarrow g^\prime_{\mu\nu}(x)= \frac{\partial x^{\prime \rho}}{\partial x^\mu}  \frac{\partial x^{\prime \sigma}}{\partial x^\mu} g_{\rho\sigma}(x^\prime(x))
\end{equation}
and use it to build the various invariants  that do not involve fermions; the latter requiring the vielbein too. 
 The inverse metric is promptly found to be
 \begin{equation}
 \label{inversmetexp}
 g^{\mu\nu}=E_a^{\phantom{a}\mu} E_b^{\phantom{b}\nu}\eta^{ab}=\eta^{\mu\nu}+\left(i\chi^\dagger \bar{\sigma}^\mu \partial^\nu \chi+ i\chi^\dagger \bar{\sigma}^\nu \partial^\mu \chi+\mathrm{h.c.}\right)+\ldots 
 \end{equation}
Note that $E_{\mu}^{\phantom{\mu}a}\eta_{ab}g^{\mu\nu}=E_b^{\phantom{b}\nu}$, meaning that we can lower and raise the various indexes with the appropriate metrics.
The determinant of the Vielbein
\begin{equation}
\det E \equiv \det E_\mu^{\phantom{\mu}a}(x)=\sqrt{-\det g_{\mu\nu}}
\end{equation}
yields the SUSY-invariant measure 
\begin{equation}
\label{invmeasure}
d^4x \det E(x) \rightarrow d^4x  | \frac{\partial x^\prime}{\partial x}| \det E(x^\prime(x))= d^4x^\prime \det E(x^\prime)\,. 
\end{equation}

The coset-ology allows us to work with objects that transform covariantly even with respect to the broken generators, upon vielbein and metric contractions. 
 These objects admit an interesting geometrical interpretation associated to the extended superspace $(x^a,\theta^\alpha_i,\theta^{\dagger\,i}_{\dot{\alpha}})$ where the 1-forms $\omega^a=dx^a-i\theta^\dagger \bar{\sigma}^ad\theta+ id\theta^\dagger \bar{\sigma}^a \theta$, $d\theta$ and $d\theta^\dagger$ are invariant under SUSY transformations $\delta_\xi(x^a,\theta,\theta^\dagger)=(- i\theta^\dagger \bar{\sigma}^a \xi +i \xi^\dagger \bar{\sigma}^a \theta, \xi,\xi^\dagger)$ \cite{Volkov:1973ix}.  These vector 1-forms are pulled-back to $E_\mu^{\phantom{\mu}a}dx^\mu$, $\partial_\mu\theta dx^\mu$ and $\partial_\mu\theta^\dagger dx^\mu$ via the map $x^\mu \rightarrow (x^a(x),\theta_{\alpha\,i}(x),\theta^{\dagger\,\alpha\,i}(x))$. These forms are nothing but the coordinates with respect to the basis $P_a$, $Q$, $Q^\dagger$ of the Maurer-Cartan form, up to identifying the Goldstino $\theta(x)=\chi(x)$ as a space-filling brane in the extended superspace. The invariant measure (\ref{invmeasure}) is nothing but the induced volume element of the Goldstino brane.
 
Every field of the strong sector that interacts with the induced metric unavoidably interacts with the Goldstini, resulting in \emph{model-independent} effects.
In addition, our construction allows higher-derivative terms that are \emph{model-dependent}, following from the direct interactions with $\nabla\chi$. We explicitly discuss these interactions in the next section. Among the model-dependent contributions, new higher-derivative geometrical objects may appear as well, such as
\begin{equation}
F_{bc}^{\phantom{bc}a}(x) \equiv E_b^{\phantom{b}\mu}(x)E_c^{\phantom{b}\nu}(x)\left( \partial_\mu E_\nu^{\phantom{\nu}a}(x)-\partial_\nu E_\mu^{\phantom{\mu}a}(x)\right)=2i\partial_b\chi^{j\dagger} \bar{\sigma}^a \partial_c\chi_j  +2i\partial_c\chi^{j\dagger}\bar{\sigma}^a \partial_b\chi_j +\ldots \, , 
\end{equation}
which transforms covariantly, $F_{bc}^{\phantom{bc}a}(x) \rightarrow  F_{bc}^{\prime\phantom{bc}a}(x^\prime(x))$ (thanks to the anti-symmetry $\mu\leftrightarrow\nu$) and represents a sort of connection $ [\nabla_a,\nabla_b]\Phi(x)=-F_{ab}^{\phantom{ab}c}\nabla_c \Phi(x)$.

\subsection{The Goldstini Effective Action}

We now have all the covariant ingredients to write an effective action, 
\begin{equation}
S_{\mathrm{SUSY}}[\chi,\Phi]=\int d^4x\, \det E\,\,  \mathcal{L}(\nabla_a\chi(x),\Phi(x),\nabla_a\Phi(x), F_{bc}^{\phantom{bc}a}(x),\ldots)=\int d^4x\, \sqrt{-\det g}\,\, \mathcal{L}
\label{Sinv}
\end{equation}
which is manifestly invariant under SUSY transformations \Eq{chitpoint} should ${\cal L}$ be Lorentz symmetric.%
\footnote{In practice this requires that all the Lorentz indexes $a$, $b$, $c$,\ldots must be saturated contracting them in pairs with the Minkowski metric $\eta_{ab}$, its inverse $\eta^{ab}$, the fully anti-symmetric $\epsilon^{abcd}$ symbol, or the sigma matrices $\sigma^{a}$. Likewise for the spinorial indexes. To construct an action that is also $R$-symmetric, the \textit{flavor} indexes $i, j, \ldots$  should be contracted among themselves with the  relevant invariant tensors, such as $\delta_{i}^{j}$. Notice that $\det E$ alone is automatically, i.e.~accidentally, invariant under maximal $R$-symmetry $U(\N)_R \sim SU(\N)_R \times U(1)_R$.} 
One can also consider nearly secluded sectors,  each enjoying its own extended SUSY simply by summing over actions of the type \Eq{Sinv}; see e.g.~\cite{Cheung:2010mc} for the case of $N$ copies of $\N=1$ SUSY. In the following we focus on a single sector with $\N$ supercharges in order to avoid the proliferation of unknown constants of possibly disparate scales.%
\footnote{However, one could in principle consider other global symmetries to reduce the unknowns. A trivial example would be semi-direct products of the type $Z_M \rtimes [\N\mbox{-SUSY}]^M$.} 

\subsubsection{Goldstini self-interactions: Akulov-Volkov}

The most important SUSY-preserving Goldstini interactions are those in \Eq{Sinv} with the least number of derivatives.  There is in fact a unique operator that yields both the leading interactions and the kinetic terms. Not surprisingly, it is the most relevant operator in a gravitational theory, the vacuum energy, injected at the SUSY phase transition, namely 
\begin{equation}
\label{CCLag}
\mathcal{L}_{CC}=-F^2 \,,
\end{equation}
which, dressed with the Goldstini, provides the extended-SUSY generalization \cite{Bardeen:1981df} of the Akulov-Volkov action \cite{Volkov:1973ix} 
\begin{equation}
\label{AVeq1}
S^{(0)}_{\mathrm{SUSY}}=\int d^4x\, - F^2 \sqrt{-\det g} = - F^2 \int d^4 x\, \det\left[\delta^a_\mu - i\chi^{j\dagger} \bar{\sigma}^a\partial_\mu\chi_j +i\partial_\mu\chi^{j\dagger}\bar{\sigma}^a \chi_j\right]\,.
\end{equation}
The scale $F$ is the SUSY-breaking scale. The action $S^{(0)}_{\mathrm{SUSY}}$ is accidentally invariant under maximal $R$-symmetry $U(\N)_R$.
 This will be important when discussing flavor violations beyond those of the SM.  
Expanding the Vielbein's determinant 
\begin{align}
\det E & =1-\left(i\chi^{j\dagger}\bar{\sigma}^\mu\partial_\mu\chi_j+\mathrm{h.c.}\right)\\
\nonumber
&+\frac{1}{2}\left[\left(i\chi^{j\dagger}\bar{\sigma}^\mu\partial_\mu\chi_j +\mathrm{h.c.}\right)^2- \left(i\chi^{j\dagger}\bar{\sigma}^a \partial_\mu\chi_j +\mathrm{h.c.}\right)\left(i\chi^{j\dagger}\bar{\sigma}^\mu \partial_a\chi_j +\mathrm{h.c.}\right)\right]+\ldots
\end{align}
we extract the kinetic term as well as the leading order interactions.
The canonically normalized Goldstini $\widetilde{\chi}$ are $\sqrt{2}F\chi=\widetilde{\chi}$. In the following we omit the tilde $\,\widetilde{}\,\,$, whenever clear.  The leading four-fermion interactions come from a dimension-8 operator that can be written as
\begin{equation}
\label{eq:S4chi}
- \int d^4x\, \frac{1}{2F^2} \left(\chi^\dagger_j \bar{\sigma}^a\partial_\mu\chi^j \right)\left(\chi^\dagger_i\bar{\sigma}^\mu\partial_a\chi^i \right)= \int d^4 x\, \frac{1}{F^2} (\chi^\dagger_i \partial_\mu \chi^\dagger_j) (\partial^\mu\chi^i \chi^j)
\end{equation}
after using the free equations of motion,  integrating by parts, and using the Fierz indentity $\partial_\nu\chi^\dagger_{\dot\beta} \bar{\sigma}^{\mu\,\dot\beta \beta} \bar{\sigma}^{\nu\,\dot\alpha \alpha} \partial_\mu\chi_\alpha=-2\partial_\mu\chi^{\dagger\,\dot\alpha}\partial^\mu \chi^\beta$. 
The Akulov-Volkov Lagrangian, with a single flavor $\mathcal{N}=1$, can be recast as $-\chi^{\dagger\,2}\square \chi^2/(4F^2)$ upon integration by parts.
This is the only operator with four fermions and two derivatives that can be built with one fermion.
With more flavors, we note that any dimension-8 operator of this kind that is consistent with the absence of more relevant interactions is necessarily selected by shift and spacetime transformations of the form \Eq{chitpoint}, which strongly suggests that they can be linked to a SUSY-breaking pattern, e.g.~for two flavors, $\N = 2$ or $(\N = 1)^2$.

\subsubsection{Model independent couplings to composite fields}

Let us consider now the case where, in addition to the Goldstini, other particles are composite, that is they emerge from the same sector that breaks SUSY spontaneously.
These fields have kinetic terms, provided by the SUSY-breaking sector itself, which need to be compensated by Goldstini insertions to make up for the absent kinetic terms of the would-be superpartners.
Since the kinetic terms can always be rescaled to a canonical form, the resulting Goldstini interactions are model independent, controlled by no free parameter except for the SUSY-breaking scale $F$ (as the Goldstini self-interactions).%
\footnote{In fact, this model independency is somewhat of a misnomer: the overall coefficient is controlled by $F$ alone because we \textit{assumed} full compositeness. Should some of these non-Goldstini states $X=\psi$, $A_\mu$ or $\phi$ be partially composite instead, the would-be model independent couplings would actually get rescaled by the degree of compositeness (squared), $|\epsilon_X|^2$, which measures how large a fraction of the kinetic term arises from the SUSY-breaking sector. Note that $\epsilon_X\neq 1$ necessarily breaks SUSY.} 

Calling $X=\psi$, $A_\mu$ or $\phi$ any composite fermion, gauge boson or scalar, we construct their SUSY-invariant kinetic terms by making the replacements $\partial_\mu X\rightarrow \nabla_a X$, $d^4x\rightarrow d^4x \sqrt{-\det g}$, $\eta^{\mu\nu}\rightarrow g^{\mu\nu}$, $A_\mu\rightarrow \mathbb{A}_a$, $F_{\mu\nu}\rightarrow \mathbb{F}_{ab}$\ldots, from which we derive the so-called model independent couplings of the composite states to the Goldstini.
They are reported in Table~\ref{tab1:indepcoup}, where we have canonically normalized the Goldstini in the last column, used the free equations of motion and integrated by parts.
The curly brackets  mean symmetrization, e.g.~$\bar{\sigma}^{\{\mu} \partial^{\nu\}} \equiv \bar{\sigma}^{\mu} \partial^{\nu}+ \bar{\sigma}^{\nu} \partial^{\mu}$. 
Notice that all these interactions respect accidentally $U(\N)_R$.

\begin{table}[t]
\begin{center}
\resizebox{\textwidth}{!}{
\begin{tabular}{|c|c|c|c|}
\hline
composites & naked terms & SUSY dressing & leading four-body interactions  \\
\hline 
$\chi$ &$ -F^2$ & $- F^2 \sqrt{-\det g}$ &   $\frac{1}{F^2} (\chi^\dagger_i \partial_\mu \chi^\dagger_j) (\partial^\mu\chi^i \chi^j)$ \\
\hline 
$\chi,\psi$ & $\frac{i}{2}\psi^\dagger_i\bar{\sigma}^\mu \partial_\mu\psi^i(x)+\mathrm{h.c.}$  & $\det E \left( \frac{i}{2}\psi^\dagger_i\bar{\sigma}^a\nabla_a\psi^i(x)+\mathrm{h.c.} \right)$  & $-\frac{1}{F^2}(\psi^\dagger_i \bar{\sigma}^a \partial_\mu \psi^i)(\chi^\dagger_j \bar{\sigma}^\mu \partial_a \chi^j)$ \\
\hline
$\chi,F$ &$-\frac{1}{4}F^A_{\mu\nu}F^{A\,\mu\nu}$ & $-\sqrt{-\det g}\, \frac{1}{4} F^A_{\mu\nu} F^A_{\rho\sigma} g^{\mu\rho}g^{\nu\sigma}$ 
&  $ -\frac{1}{4F^2}F^A_{\mu\nu}F^{A\,\mu}_{\phantom{\mu}\rho}\left(i \chi^\dagger_i \bar{\sigma}^{\{\rho}\partial^{\nu\}}\chi^i+\mathrm{h.c.}\right)$  \\
\hline
$\chi,\phi$ &$\partial_\mu \phi^{i\dagger} \partial^\mu \phi_i$ & $\sqrt{-\det g}\, g^{\mu\nu}\partial_\mu \phi^{i\dagger} \partial_\nu \phi_i $ &  $\frac{1}{2F^2}\left(i\chi^\dagger_j \bar{\sigma}^{\{\mu} \partial^{\nu\}} \chi^j+\mathrm{h.c.}\right)\partial_\mu\phi^{i\dagger} \partial_\nu\phi_i$\\
\hline
\end{tabular}}
\end{center}
\caption{\footnotesize\emph{Model independent couplings for composite particles $X=\psi,A_\mu, \phi$. Partially composite particles have these couplings weighted by their degree of compositeness squared $|\epsilon_X|^2$. If the scalar $\phi$ is a composite NGB one should dress the whole kinetic term built with the $\mathcal{D}$-symbol} $\mathcal{D}^\mu_i = \partial^\mu \phi_i + O(\phi^3)$.}
\label{tab1:indepcoup}
\end{table}

When considering the couplings of the strong sector to external (gauge) sources, an important r\^ole is played in our construction by the $R$-symmetry current, the conserved Noether current associated with infinitesimal $R$-symmetry transformations $\chi\rightarrow \chi - i\omega_A T^A \chi$  ($T^A$ are the appropriate $R$-generators).
This current can be found by noticing that in our geometrical construction Goldstini enter the action only through geometrical objects, so that the $R$-current is directly related to the energy-momentum tensor (associated to the model-independent universal contributions)
\begin{equation}
\label{Tminimal}
T_a^{\phantom{a}\mu}=-\delta S/\delta E_\mu^{\phantom{\mu}a}=-\mathcal{L}E_a^{\phantom{a}\mu}+\det E\left[\left(\frac{i}{2}\psi^{i\dagger} \bar{\sigma}^b E_b^{\phantom{b}\mu} \nabla_a\psi_i +\nabla_a\phi^{i\dagger} \partial_\nu\phi_i g^{\mu\nu}+\mathrm{h.c.}\right)-\mathbb{F}_{ab}\mathbb{F}^{cb}E_c^{\phantom{c}\mu}\right]\,.
\end{equation}
From this we find, for the $R$-current \cite{Clark:2000rv}, 
\begin{equation}
\label{R-current}
R^{A\,\mu}=\frac{1}{F^2}T_{a}^{\phantom{a}\mu}\chi^\dagger \bar{\sigma}^aT^A\chi=\left(\chi^\dagger \bar{\sigma}^aT^A\chi\right)\left(\delta^{\mu}_a+\frac{i}{2F^2}\chi^{j\dagger} \bar{\sigma}^\mu\overleftrightarrow\partial_a\chi_j +\ldots\right) \, .
\end{equation}
Analogously, the SUSY currents can also be expressed in terms of the energy-momentum tensor, namely   $S^{\mu\,\alpha\, j}= \sqrt{2}T_{a}^{\phantom{a}\mu} \chi^{j\dagger}_{\dot\beta}(\bar{\sigma}^{a})^{\dot\beta\alpha }/F$ and  $S^{\dagger \mu\,\dot\alpha}_j= \sqrt{2} T_{a}^{\phantom{a}\mu} (\bar{\sigma}^{a})^{\dot\alpha\beta} \chi_{j\beta}/F$, with  $j=1,\ldots, \N$.

\subsubsection{Model dependent couplings to composite fields}

The model dependent couplings may or may not be there, depending on the details of the UV theory.
Generically they do not respect maximal $R$-symmetry, but just some of its subgroups.

In Table~\ref{tab2:modedepcoup} we show a few illustrative examples assuming that the $\chi_i$ carry at least a $U(1)_R$, i.e.~$\chi_i\rightarrow \chi_i e^{i\alpha_R}$, that the scalar $\pi$ is naturally light because it is a Nambu-Goldstone boson (NGB), and we restrict to a singlet fermion.
In the last row we kept the leading term in the number of NGBs of the $\mathcal{E}$-symbol, $\mathcal{E}_b^A = -i\pi^\dagger T^A \overleftrightarrow \nabla_b \pi + O(\pi^4)$.
We also enforced maximal $R$-symmetry, $G_R = U(\N)_R$, for simplicity. This can also be done for the first and second rows by choosing $c^j_i, d^j_i \propto \delta^j_i$. Alternatively, choosing e.g.~$c^j_i, d^j_i=\mathrm{diag}(c^{(N)} \mathbb{I}_{N\times N}, c^{(M)} \mathbb{I}_{M\times M})$ with $\N=N+M$ corresponds to realize linearly only the $SU(N)_R\times SU(M)_R\times U(1)_R$ subgroup of $U(\N)_R$ under which $\chi=(\mathbf{N},1)_x \oplus (1,\mathbf{M})_{-x N/M}$\,.
The size of these interactions depends on the assumptions about the UV theory, e.g.~the size of couplings with additional heavy states.

\begin{table}[t]
\begin{center}
\resizebox{\textwidth}{!}{
\begin{tabular}{|c|c|c|}
\hline
$G_R$ &  SUSY Lagrangian  & leading interactions  \\
 \hline
$\psi=\mathbf{1}$ & $c^j_i \det E\, (\nabla_a \chi^{i\dagger}  \bar{\sigma}^b \nabla^a \chi_j)(\psi^\dagger \bar{\sigma}_b \psi)$ &  $c^j_i\frac{1}{F^2} (\partial_\nu \chi^{i\dagger}  \bar{\sigma}^\mu \partial^\nu \chi_j)(\psi^\dagger \bar{\sigma}_\mu \psi)$\\
\hline
$\pi=\mathbf{1}$, NGB & $d^j_i\det E\, \left(\nabla_a \chi^{i\dagger} \bar{\sigma}^b \nabla^a \chi_j\right) \nabla_b \pi$ & five-body or $\propto m_\pi, m_\chi$\\
\hline
$\pi=\Yfund$, NGB & $ c \det E\,\left(\nabla_a \chi^\dagger  \bar{\sigma}^b T^A \nabla^a \chi\right) \left(i\pi^\dagger T^A  \overleftrightarrow\nabla_b \pi\right) $ & $c \frac{1}{F^2} \left(\partial_\mu \chi^\dagger  \bar{\sigma}^\nu T^A \partial^\mu \chi\right) \left(i\pi^\dagger T^A  \overleftrightarrow\partial_\nu \pi\right)$ \\
\hline
\end{tabular}}
\end{center}
\caption{\footnotesize\emph{Model dependent couplings. For composite particles $c^j_i, d^j_i, c = O(1)$.}}
\label{tab2:modedepcoup}
\end{table}

The generalization to non-NGB scalars, non-singlet fermions, or other choices of conserved $R$-symmetries and its representations is straightforward. 
For example, a scalar without a shift symmetry has a model independent coupling proportional to its mass squared, $-m_{\phi}^2|\phi|^2 \det E$, which contributes to a 6-body scattering.%
\footnote{One could naively think that $|\phi|^2$ contributes to the $\chi\chi\rightarrow \chi\chi$ elastic Goldstino scattering when $\phi$ gets a VEV, and that there could be a lower bound on how much tachyonic the mass squared can get, from the usual positivity of the Goldstino amplitude \cite{Bellazzini:2016xrt}.  However, a SUSY-preserving potential for $\phi$ must have vanishing vacuum energy, e.g.~$V=\lambda(|\phi|^2-f^2/2)^2$, since by construction we include it all in the definition of $F$ in \Eq{CCLag}. This ensures that the $\chi\chi\rightarrow \chi\chi$ is actually not affected as long as the potential is SUSY-preserving.}
Higher-derivative non-NGB couplings would be of the form $\det E\, (\nabla_a\chi^\dagger \bar{\sigma}^b\overleftrightarrow\nabla_b \nabla^a \chi)|\phi|^2$, with a model dependent coefficient.

Since the model dependent contributions include extra Goldstini beyond those contained in $\det E$ or the metric, the resulting equations of motion and the conserved $R$-current are different than those originating from the model independent contributions.

\subsection{Embedding Quarks and Leptons}
\label{sec:emb}

We now take a step towards the SM and identify (some of) the SM fermions as Goldstini filling $R$-symmetry multiplets. We still work in the limit where gauge and Yukawa interactions vanish, but ensure that the strong sector itself respects the symmetries of the SM, i.e.~gauge and flavor symmetries, $G_{SM}\equiv G_{Gauge}\times G_{Flav}$, with $G_{Gauge}=SU(3)_C\times SU(2)_L\times U(1)_Y$ and $G_{Flav}= SU(3)_{Flav}^5\times U(1)_B\times U(1)_L$.%
\footnote{Whenever including the right-handed neutrinos $\nu^c$ we actually consider $SU(3)_{Flav}^6$.}
Since the only symmetry under which Goldstini transform is by definition the $R$-symmetry, $G_R$, the embedding of the SM fermions is possible only if their gauge and flavor symmetries can be embedded into the $R$-symmetry, $G_{SM}\subset G_R$.

\subsubsection{Maximal $R$-symmetry}

$N$ free Weyl fermions, which is what the SM reduces to in the limit $g_{SM}\equiv \{g,g^\prime,g_s,Y_\psi\}\to 0$, enjoys an $U(N)$ symmetry such that they transform in the fundamental representation.
While gauge interactions in the SM break explicitly this $U(45)$ symmetry down to the smaller $G_{SM}$, the actual representations of quarks and leptons with respect to $G_{SM}$ do not fit a fundamental representation of the original $U(45)$, simply because the $SU(2)_L$ singlet and doublet quarks come in different color representations, namely a $\mathbf{3}^*$ and $\mathbf{3}$ respectively.

These facts may be replicated in the strong sector; the lowest-dimensional interactions (the Akulov-Volkov Lagrangian \Eq{AVeq1}) are accidentally $U(\N)_R$ symmetric, with Goldstini in the fundamental representation. 
Adding higher-dimensional terms, e.g.~$c_{ij}^{kl} \nabla_a \chi^i \nabla^a \chi^j\nabla_b \chi^{\dagger}_{k} \nabla^b \chi^{\dagger}_{l}$, generically breaks $U(\N)_R$ to a subgroup under which the Goldstini transform in various representations that no longer fit, generically, into the fundamental of $U(\N)_R$.
However, if SM matter is to be identified as Goldstini, these representations must fit at the very least the proper SM representations of $G_{Gauge}$.
While it is technically possible that $G_{R}=G_{gauge}$ (or $G_{R}=G_{SM}$) is strictly smaller than $U(\N)$ and that the Goldstini representations are exactly those of the SM fermions, it would be very surprising, lacking a dynamical reason.
With no better option, we assume instead in the following that the SUSY-breaking sector is maximally $R$-symmetric, i.e.~$G_{R}=U(\N)_R$. The reduction to $G_{SM}$ and then to $G_{Gauge}$  is entirely due to the external SM couplings $g_{SM}$, rather than the SUSY-preserving parameters of the strong sector.  In other words, we extend the paradigm of MFV from $G_{SM}$ to the largest group that can be simultaneously preserved by the strong sector and the free theory, i.e.~$U(\N)_R$. 

Interestingly, in most cases the assumption of maximal $R$-symmetry does not obstruct a proper embedding of the SM fermions as Goldstini.
Maximal $R$-symmetry is often obtained when the strong sector has the least number of supercharges compatible with the assigned Goldstini content, see Table~\ref{tab:embeddings}. 
In other words, the representations of the SM fermions often fit in the fundamental of $U(\N)_R$.
Assuming maximal $R$-symmetry is relevant, for what regards the SM embedding, only when both doublet and singlet quarks are Goldstini, 
since it requires a number of supercharges larger than the number of Goldstini-like SM fermions, implying extra light fermions in the spectrum.

Maximal $R$-symmetry means, in practice, that we should embed the SM matter representations that are associated to Goldstini in the fundamental of $U(\N)_R$. To do so, the following decompositions turns out to be useful%
\footnote{In the first case, an index $I$ in the fundamental that runs from $1$ to $N\cdot M$ can be split into a collective pair of indexes $I=(i,j)$  where $i=1,\ldots N$ and $j=1,\ldots M$. In the second case the collective index $I=1,\ldots N+M$ is split in two separate indexes, $i=1,\ldots, N$ and $j=N+1,\ldots,N+M$.}
\begin{align}
\label{group1}
& SU(N\times M)\supset SU(N)\times SU(M) \hspace{-1.2cm} &\Rightarrow& \quad \Yfund=(\Yfund,\Yfund) \, ,\\
& SU(N+M)\supset SU(N)\times SU(M)\times U(1) \hspace{-1.2cm} &\Rightarrow& \quad \Yfund=(\Yfund,\mathbf{1})_{a/N} \oplus (\mathbf{1},\Yfund)_{-a/M} \, .
\label{group2}
\end{align}

\subsubsection{Embedding Leptons}

Let us start with the simplest case: the singlet electron, charged only under hypercharge, $e^c=(\mathbf{1},\mathbf{1})_1$ under $SU(3)_C\times SU(2)_L\times U(1)_Y$. In this case we promptly identify the hypercharge with the $U(1)_R$ of $\N=1$, that is $\chi=e^c$.

The case where all three generations of singlet electrons are Goldstini is slightly more interesting as we insist on a $SU(3)_{e^c}$ flavor symmetry. The proper embedding is via a triplet $\pmb{e}^{c}=(e^c, \mu^c,\tau^c) = \mathbf{3}_1$ of $SU(3)_R \times U(1)_R$ in $\N=3$, where the $U(1)_R$ and $SU(3)_R$ factors are identified with the hypercharge and the flavor group respectively. In this case the $R$-symmetry index $j$ is nothing but that the flavor index, $\chi_j=e^{c}_{j}$. 

Embedding only one lepton doublet, $\ell=(\mathbf{1},\mathbf{2})_{-1/2}$ under $G_{Gauge}$, requires $\N=2$ and again maximal $R$-symmetry $U(2)_R\sim SU(2)_L \times U(1)_Y$. 
The $R$-symmetry index is an electroweak index in this case, $\ell_{j}=(\nu_L, e_L)_j=\chi_j$. 
Similarly, including all lepton doublets requires to consider $\N=6$ and to decompose the fundamental of $U(6)_R$ as $\mathbf{6}_{-1/2}=(\mathbf{2},\mathbf{3})_{-1/2}$ with respect to $SU(2)_R \times SU(3)_R \times U(1)_R$. One then identifies $SU(2)_R \times U(1)_R$ with $SU(2)_L \times U(1)_Y$ and $SU(3)_R$ with the flavor group. 
  
With little extra effort we can embed all leptons, including the singlet neutrinos $\nu^c$, taking $\N=12$  and the maximal $R$-symmetry group $U(12)_R$. Such a large $R$-symmetry contains the proper subgroups $SU(12)_R \times U(1)_R \supset SU(6) \times SU(6) \times U(1)_R \times U(1)_A$ that in turn $\supset \left(SU(3) \times SU(2)\right) \times \left(SU(3)\times SU(3) \times U(1)_C\right) \times U(1)_R \times U(1)_A$, where the three $SU(3)$'s are identified with the flavor groups acting on the flavor triplets $\pmb{\ell}$, $\pmb{e}^c$ and $\pmb{\nu}^c$,  while the $SU(2)$ factor is identified with $SU(2)_L$. 
The last abelian $U(1)_C$ allows us to give independent hypercharges to doublet and singlet leptons.
Explicitly, the fundamental of $U(12)_R$ decomposes as 
 \begin{align}
 \mathbf{12}_r & =(\mathbf{6},\mathbf{1})_{r,a}\oplus (\mathbf{1},\mathbf{6})_{r,-a}=(\mathbf{3},\mathbf{2},\mathbf{1},\mathbf{1})_{r,a,0}\oplus (\mathbf{1},\mathbf{1},\mathbf{3},\mathbf{1})_{r,-a, c}\oplus (\mathbf{1},\mathbf{1},\mathbf{1},\mathbf{3})_{r,-a,-c}
 \nonumber 
 = \pmb{L}\oplus \pmb{e}^c \oplus \pmb{\nu}^c
 \end{align}
under the chain of subgroups we have mentioned above.
The hypercharge is identified with $Y=-A/(2a)+C/(2c)$, the lepton number is $L=A/a$, while the $U(1)_R$ plays no role (we could have demanded just $SU(12)_R$ rather than the maximal $U(12)_R$).
We summarize these and other  cases in Table~\ref{tab:embeddings}. 

\subsubsection{Embedding Quarks and Extra Exotics}

The embedding of either $SU(2)_L$ doublet or singlet quarks works like for leptons, as we show in Table~\ref{tab:embeddings}.
Things become more complicated when we embed the quark doublets together with the singlets inside the same fundamental representation of $SU(\N)$. This is due to the difficulty in obtaining both a $\mathbf{3}^*$ (for $d^c$ and $u^c$) and a $\mathbf{3}$ (for $q$) of $SU(3)_C$ when decomposing the fundamental of $SU(\N)$.%
\footnote{The $\mathbf{36}$ is the minimal representation that could in principle accommodate the $18+9+9$ quarks $\pmb{q}$, $\pmb{u}^c$ and $\pmb{d}^c$. However, the decomposition of $U(36)_R$ into $\left(SU(2)\times [SU(3)]^2\right)\times \left([SU(3)]^2 \times [SU(3)]^2\right) \times U(1)_R\times U(1)_A \times U(1)_C$ is  $\mathbf{36}_r= (\mathbf{2},\mathbf{3},\mathbf{3},\mathbf{1},\mathbf{1},\mathbf{1},\mathbf{1})_{r,a} \oplus (\mathbf{1},\mathbf{1},\mathbf{1},\mathbf{3},\mathbf{3},\mathbf{1},\mathbf{1})_{r,-a,c} \oplus (\mathbf{1},\mathbf{1},\mathbf{1},\mathbf{1},\mathbf{1},\mathbf{3},\mathbf{3})_{r,-a,-c}$ which does not contain a color $\mathbf{3}^*$ with the other three $SU(3)$'s identified as the flavor groups.}
The solution to this problem is to add extra states to fill a larger multiplet that can give rise to $\mathbf{3}^*$'s, since a $\mathbf{3}^*$ of $SU(3)$ can be built out of two fundamentals, $\mathbf{3}\otimes \mathbf{3}=\mathbf{3}^* \oplus \mathbf{6}$. Then, the smallest group that can accommodate all quarks (all flavors) is $SU(72)_R \times U(1)_R$, with the $\mathbf{72}_r$ decomposing as 
\begin{align}
\mathbf{72}_r   = & (\mathbf{2},\mathbf{3},\mathbf{3},\mathbf{1},\mathbf{1},\mathbf{1},\mathbf{1})_{r,3a,0} 
\label{eqof72}
 \oplus (\mathbf{1},\mathbf{1},\mathbf{1},\mathbf{3},\mathbf{3}^*,\mathbf{1},\mathbf{1})_{r,-a,c}  \oplus (\mathbf{1},\mathbf{1},\mathbf{1},\mathbf{3},\mathbf{6},\mathbf{1},\mathbf{1})_{r,-a,c} \\
\nonumber
 & \oplus (\mathbf{1},\mathbf{1},\mathbf{1},\mathbf{1},\mathbf{1},\mathbf{3},\mathbf{3}^*)_{r,-a,-c} \oplus (\mathbf{1},\mathbf{1},\mathbf{1},\mathbf{1},\mathbf{1},\mathbf{3},\mathbf{6})_{r,-a,-c}
= 
\pmb{q}\oplus
\pmb{u}^c \oplus
\pmb{X}_{-2/3}\oplus
\pmb{d}^c \oplus
\pmb{X}_{1/3}
\end{align}
with respect to the subgroup 
\begin{equation}
\left(SU(2)\times [SU(3)]^2\right)\times \left( [SU(3)]^2 \times [SU(3)]^2\right) \times U(1)_R\times U(1)_A \times  U(1)_C \,. \nonumber
\end{equation}  
A diagonal $SU(3)$ out of three $SU(3)$'s is identified with the color group while three other $SU(3)$'s represent the flavor group $SU(3)_{q}^{Flav}\times SU(3)^{Flav}_{d}\times SU(3)_{u}^{Flav}$ (this matter content is then consistent with two extra global $SU(3)$ factors). The hypercharge reads $Y=-R/(12r)+A/(12a)-C/(2c)$, whereas the baryon number is $B=-R/(6r)+A/(6a)$.

\begin{table}[t]
\begin{center}
\resizebox{\textwidth}{!}{
\begin{tabular}{|c|c|c|}
\hline
Goldstini& $G_{Gauge}\times G_{Flav}$& ${\cal N}_{min}$\\
\hline\hline
$e^c$ & $U(1)_Y$ & ${\cal N}=1$ \\
\hline
$\ell_e $ & $SU(2)_L \times U(1)_Y$ & ${\cal N}=2$ \\
\hline
$\ell_e $, $e^c$ & $SU(2)_L  \times U(1)_Y \times U(1)_{L_e} $ & ${\cal N}=3$ \\
\hline
$\ell_e$, $e^c$, $\nu_e^c$ & $SU(2)_L \times U(1)_Y \times U(1)_{L_e}$ & ${\cal N}=4^*$ \\
\hline
$d^c$ or $u^c $ &$SU(3)_C \times U(1)_Y$ & ${\cal N}=3$ \\
\hline
$\pmb{e}^c $ & $U(1)_Y\times SU(3)^{Flav}_{e}$ & ${\cal N}=3$ \\
\hline
$\pmb{\ell}$ & $SU(2)_L\times U(1)_Y \times SU(3)^{Flav}_{\ell}$ & ${\cal N}=6$ \\
\hline
$\pmb{\ell}$, $\pmb{e}^c$ & $SU(2)_L\times U(1)_Y \times U(1)_{L} \times SU(3)_{Flav}^{2}$ & ${\cal N}=9$ \\
\hline
$\pmb{\ell},\pmb{e}^c,\pmb{\nu}^c$ & $SU(2)_L\times U(1)_Y\times U(1)_L \times  SU(3)_{Flav}^3$ & ${\cal N}=12^*$ \\
\hline
$\pmb{d}^c$ or $\pmb{u}^c$&$SU(3)_C\times U(1)_Y\times SU(3)_{d(u)}^{Flav}$& $\N=9$\\
\hline
$\pmb{q}$& $SU(2)_L \times SU(3)_C \times U(1)_Y \times SU(3)_{q}^{Flav}$& $\N=18$\\
\hline
$\pmb{d}^c$, $\pmb{u}^c$ & $SU(2)_L\times [SU(3)_C]^2\times [U(1)_Y]^2 \times SU(3)_{Flav}^2$& $\N=18$\\
\hline
$\pmb{q},\pmb{d}^c,\pmb{u}^c,\pmb{X}_{-2/3,1/3}$ & $ SU(2)_L \times [SU(3)_C]^3 \times [U(1)_Y]^2 \times U(1)_B \times SU(3)_{Flav}^3$& $\N=72\, (36)$\\
\hline
$\pmb{\ell},\pmb{e}^c,\pmb{\nu}^c,\pmb{q},\pmb{d}^c,\pmb{u}^c,\pmb{X}_{-2/3,1/3}$ & $ [SU(2)_L]^2 \times [SU(3)_C]^3 \times [U(1)_Y]^4\times U(1)_{B} \times U(1)_L \times SU(3)_{Flav}^6$& $\N=84\, (48)$\\
\hline
\end{tabular}}
\end{center}
\caption{\footnotesize\emph{The minimal supersymmetry ${\cal N}_{min}$ needed to reproduce the correct quantum numbers of SM fields under gauge and flavor symmetries~$G_{Gauge}\times G_{Flav}$, including the baryon and lepton numbers $U(1)_B$ and $U(1)_L$. Boldface characters denote flavor triplets. The notation $[SU(m)_{C,L,Y}]^n$ means there are $n$ $SU(m)$ factors of which only one linear combination corresponds to the gauged $SU(m)_{C,L,Y}$ group. The asterisk $^*$ marks whether $SU(\N)_R$, as opposed to $U(\N)_R$, is required. The number of supercharges inside parenthesis refers to $\N_{min}$ for non-maximally $R$-symmetries (with Goldstini not in the fundamental representation).}}
\label{tab:embeddings}
\end{table}

The prediction of maximal $R$-symmetry is then that there are extra (pseudo-)Goldstini $X_{-2/3,1/3}$ that are colored and charged under $Y$, transforming as 
\begin{equation}
\label{sixpletx}
\pmb{X}_{-2/3,1/3}=(\mathbf{6},\mathbf{1})_{-2/3,1/3}
\end{equation}
under $G_{Gauge}$. From the decomposition in \Eq{eqof72} we see that these exotic states (aka quixes) are also triplets of the flavor groups, either $SU(3)_{u}$ or $SU(3)_{d}$, depending on their hypercharge. 

Similarly, embedding \emph{all} SM quarks and leptons into Goldstini requires $\N=84$ supercharges, where the fundamental splits as $\mathbf{84}=\mathbf{12}\oplus \mathbf{72}$, and we can apply the results derived above.
The extra $36$ states correspond again to three families of the two exotic color sextets.

There are two interesting points related to this extra matter content. 
First, they give rise to $[SU(3)_C]^2 U(1)_Y$ and $[U(1)_Y]^3$ anomalies. In order to cancel them there must exist extra light colored states, such as complex conjugate color sextets, 
\begin{equation}
\label{sixplety}
\pmb{Y}_{2/3,-1/3}=(\mathbf{6}^*,\mathbf{1})_{2/3,-1/3} \, , 
\end{equation}
to form, along with $X$, real representations of $G_{Gauge}$. The $\pmb{Y}$'s are anti-triplets of flavor.
We will loosely refer to these Dirac fermions $\Psi_{\mathbf{6}}=X\oplus Y^c$ as the sextet $\mathbf{6}$.
The sextet is light, relative to the strong coupling scale $m_* \sim \sqrt{g_* F}$, since a mass term $m_{\mathbf{6}} \, X Y$ requires breaking SUSY explicitly (see section~\ref{sec:break}).
Second, the sextet contributes significantly to the running of the strong coupling: four extra Weyl fermions (two pseudo-Goldstini $X$ and two extra fields $Y$), in the $\mathbf{6}$ and $\mathbf{6}^*$ of $SU(3)_C$, per three generations, imply a contribution $\delta b_{\mathbf{6}}=-20$ to the 1-loop $\beta$-function $\beta_{1-loop}= -b g^3/(16\pi^2)$.
This implies that above $m_{\mathbf{6}}$ the $\beta$-function changes sign and would hit a Landau-pole at roughly $\Lambda\simeq m_\mathbf{6} \mathrm{Exp}[2\pi/(b \alpha_s(m_\mathbf{6}))] \approx 10^2 m_\mathbf{6}$, where we have taken $\alpha_s(m_\mathbf{6}) \approx 0.09$.
This sets an upper bound on the scale of strong coupling $m_*$, which should enter before $\Lambda$. Equivalently, this sets a lower bound on $m_{\mathbf{6}}/m_* \gtrsim 10^{-2}$.

\section{Explicit SUSY Breaking}
\label{sec:break}

So far we have discussed the properties that characterize the strongly interacting sector, in different cases where one or more SM fermions are Goldstini, but in the limit of vanishing SM interactions $g_{SM} \to 0$. The leading interactions genuine of goldstino-compositeness are controlled by dimension-8 operators, which are large at high energy, but vanishingly small at small energy.
These interactions preserve a maximal $R$-symmetry, which contains the relevant global groups to be gauged and flavor groups.

This picture is necessarily distorted at least by the marginal SM interactions, which are small but become leading at sufficiently small energy. These interactions break SUSY explicitly, the Goldstino-like SM fermions become pseudo-Goldstini, and new effects are generated, which we estimate in this section.

\begin{table}[t]
\begin{center}
\begin{tabular}{|l|l|}
\hline
\hspace{0.5cm} {\bf Fermion-Gauge} & \hspace{1.25cm} {\bf Fermion-Higgs} \\
$\Op^{\psi }_{B}=D^\nu B_{\mu\nu}( \bar \psi_{L,R}\gamma^\mu \psi_{L,R})$ &$\Op^\psi_{L,R}=(i H^\dagger {\lra {D}_\mu} H)( \bar \psi_{L,R}\gamma^\mu \psi_{L,R})$ \\ 
$\Op^{\psi }_{W}= D^\nu W_{\mu\nu}^a( \bar \psi_L\sigma^a\gamma^\mu \psi_L)$&$\Op^{(3)\, \psi}_{L}=(i H^\dagger \sigma^a {\lra {D}_\mu} H)( \bar \psi_L\sigma^a\gamma^\mu \psi_L)$\\
\cline{1-1}
\hspace{1.25cm} {\bf Dipoles} & $\Op_{y_\psi}=|H|^2 \bar \psi_L H \psi_R$\\
\cline{2-2}
$\Op_{DB} = \bar \psi_L \sigma^{\mu \nu} H \psi_R B_{\mu \nu} $ & \hspace{1.25cm} {\bf Four-Fermions}\\
$\Op_{DW} = \bar \psi_L \sigma^{\mu \nu} \sigma^a H \psi_R W^a_{\mu \nu} $&$\Op_{4\psi}=\bar\psi\gamma_\mu\psi\bar\psi\gamma^\mu\psi$\\
$\Op_{DG} =  \bar \psi_L \sigma^{\mu \nu} H T^A \psi_R G^A_{\mu \nu}$&\\
\hline
\end{tabular}
\end{center}
	\caption{\footnotesize\emph{Dimension-6 operators involving fermions relevant for our analysis.}}
\label{d6ops}
\end{table}

The SM interactions should be thought as spurions of SUSY breaking,
\beq\label{expbreak}
\Delta \Lag_{break} = - Y_{\chi\, ij} \left[ \chi^i \chi^j H + \dots \right] + \mathrm{h.c.} + g_V \left[ V_\mu^A R^{A \mu} + \dots \right]\,.
\eeq
The Yukawas $Y$ also break the flavor symmetries and we assume these are the only sources of flavor breaking, thus realizing the MFV paradigm~\cite{DAmbrosio:2002vsn}.
Gauge interactions arise through weakly gauging some of the $R$-currents  $R^{A \mu}$, \Eq{R-current}. 
The dots in \Eq{expbreak} represent the generalization of the minimal symmetry-breaking interactions (the first terms in brackets) by  operators with the same field, symmetry, and spurionic content but with extra (covariant) derivatives, suppressed by $m_*$.

We focus first on the scenario where the only light composite d.o.f.'s are the pseudo-Goldstini, while non-Goldstini SM fermions, the Higgs and the gauge bosons are elementary; we comment below on extensions of this picture.
In such minimal scenario no dimension-6 operators are generated directly by the strong dynamics, but some might be generated via loops involving both the strong dynamics and the spurions in \Eq{expbreak} -- a list of the interesting dimension-6 operators is reported in Table~\ref{d6ops}.
We estimate these effects by simple power counting, derived from the leading loop diagrams, some of which are shown in \Fig{fig:diags}.

\subsubsection*{$Z$ couplings}

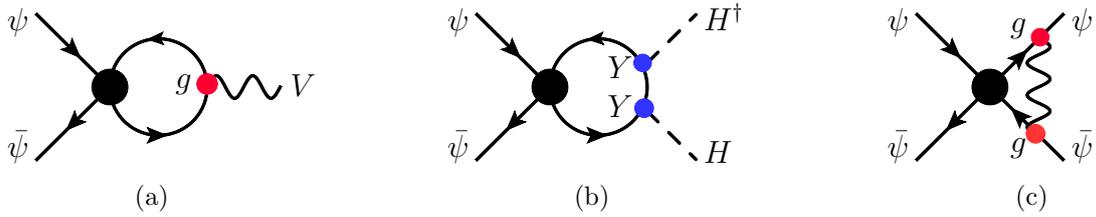
\begin{figure}[tp]
\centering
\hspace{1cm}\subfigure[]{\begin{picture}(100,80)
\put(0,0){ \includegraphics[height= 2cm]{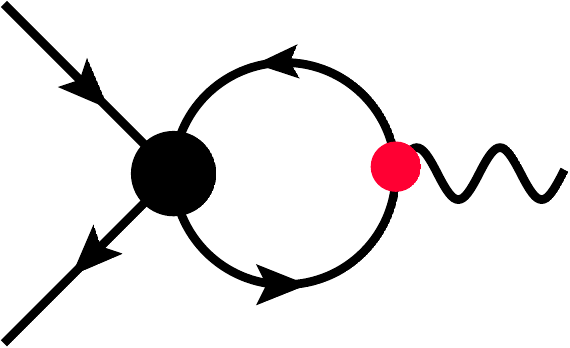}}
\put(-6,50){$\psi$}
\put(-6,2){$\bar\psi$}
\put(57,28){$g$}
\put(100,25){$V$}
\end{picture}\label{figa}}
\hspace{2cm}
\subfigure[]{\begin{picture}(100,80)
\put(0,0){ \includegraphics[height= 2cm]{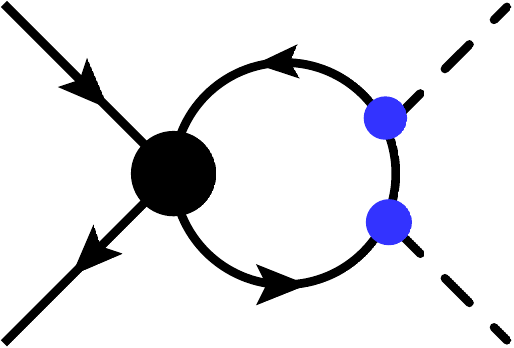}}
\put(-6,50){$\psi$}
\put(-6,2){$\bar\psi$}
\put(54,32){$Y$}
\put(54,17){$Y$}
\put(90,50){$H^\dagger$}
\put(90,0){$H$}
\end{picture}\label{figb}}
\hspace{2cm}
\subfigure[]{\begin{picture}(100,80)
\put(0,0){ \includegraphics[height= 2cm]{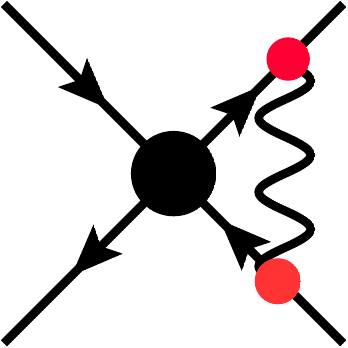}}
\put(-6,50){$\psi$}
\put(-6,2){$\bar\psi$}
\put(63,50){$\psi$}
\put(63,2){$\bar\psi$}
\put(40,49){$g$}
\put(40,4){$g$}
\end{picture}\label{figc}}
\caption{\footnotesize \emph{Some leading diagrams contributing to dimension-6 operators with fermions. Black blobs denote strong interactions, while little red (blue) ones denote SM gauge (Yukawa) couplings that break SUSY explicitly.}}
\label{fig:diags}
\end{figure}

The largest effects involve operators with external fermions and the least number of SUSY-breaking spurions. In particular, loops of pure strong sector with a gauge boson insertion generate $\Op^{\psi }_{W,B}$ (see e.g.~\Fig{figa}) with coefficients
\begin{equation}\label{cwb}
c^{\psi}_{W,B} \sim \frac{g}{m_*^2}\,.
\end{equation}
Through a field redefinition (which corresponds to the equations of motion) these operators are equivalent to combinations of $\Op^\psi_{L,R}$,  $\Op^{(3)\, \psi}_{L}$ and $\Op_{4\psi}$, with coefficients $c^{\psi}_{L,R}\sim c^{(3)\psi}_{L}\sim g^2/m_*^2$ and $c_{4\psi}\sim g^2/m_*^2$.
The former modify the couplings of the $Z$ boson to fermions, which are constrained at the permille level from measurements at LEP-1.
Consequently, goldstino-compositeness of doublet leptons (or $L$-handed, $\ell_L$, in Dirac notation), is constrained as\cite{Pomarol:2013zra,Ellis:2014jta}
\beq
\pmb{\ell} \textrm{ Goldstini:} \quad m_* \gtrsim 2.5 \TeV \,,
\eeq
while goldstino-compositeness of singlet  leptons (or $R$-handed, $e_R$), even if probed by the same data, gets a milder constraint due to the $(g'/g)^2$ suppression of the corresponding operators
\beq
\pmb{e}^c \textrm{ Goldstini:} \quad m_* \gtrsim 2 \TeV \,.
\eeq
We will see in section~\ref{sec:LEP} that, because of the low energies accessible at LEP, the constraints on pseudo-Goldstini leptons from their defining dimension-8 operators  are weaker than the indirect ones derived here.

Given the constraints on modifications of the $L$-handed quark couplings to gauge bosons, for instance from LEP measurements of the $Z$ decay to hadrons, their goldstino-compositeness is indirectly constrained at the same level as that of $L$-handed leptons
\beq
\pmb{q} \textrm{ Goldstini:} \quad m_* \gtrsim 2.5 \TeV \,.
\label{qewpt}
\eeq
On the contrary, goldstino-compositeness of $R$-handed quarks is not very constrained via the effects of $\Op_R^{u, \, d}$, given that the sensitivity of LEP to non-standard $R$-handed quark couplings is significantly reduced by their small SM coupling to the $Z$.

The only relevant Yukawa-induced operators are those associated to up quarks as pseudo-Goldstini, because of the large top Yukawa. In particular, if both $L$- and $R$-handed tops are pseudo-Goldstini, we find (from e.g.~\Fig{figb}) an additional contribution to
\begin{equation}
c^{(3)\psi}_{L} \sim c^{\psi}_{L,R}\sim  \frac{Y_t^2}{m_*^2}\,.
\label{cLtop}
\end{equation}
This could be the dominant effect, since it modifies the $Zb\bar b$ coupling, implying
\beq
q_3 \textrm{ and } t^c \textrm{ Goldstini:} \quad m_* \gtrsim 5 \TeV\,.
\label{topZbb}
\eeq
However, if $q_3$ is elementary this bound disappears, while if $t^c$ is elementary, loop diagrams like \Fig{figb} become weak and thus suppressed by $g_*^2/16\pi^2$, implying $c^{(3)\psi}_{L} \sim c^{\psi}_{L} \sim Y_t^2g_*^2/16\pi^2m_*^2$, which might alleviate the bound if the strong coupling is not maximal, $g_* < 4 \pi$.
Other constraints associated with \Eq{cLtop} come from top physics only and therefore are mild.

\subsubsection*{Higgs couplings}

The operators $\Op_{y_\psi}$ are generated with coefficients $c_{y_\psi}\sim Y_\psi^3/m_*^2$ (or $Y_\psi g^2 g_*^2/16 \pi^2 m_*^2$) given that the discrete symmetry $H\rightarrow -H$ is broken only by the Yukawas. The experimental constraints from measurements of the Higgs couplings to fermions are not competitive enough to make these effects relevant. Furthermore, this estimate is reduced by $g_*^2/16\pi^2$ if only the $L$- or $R$-handed fermions are pseudo-Goldstini.

\subsubsection*{Four-fermion contact interactions}

The operators $\Op_{4\psi}$ can be generated with coefficients $c_{4\psi}\sim g^2/m_*^2$ from the field redefinition described below \Eq{cwb}, or directly from diagrams like \Fig{figc}, carrying an additional $g_*^2/16\pi^2$ suppression due to the elementary vector boson in the loop. These operators contribute to the same observables as the bona fide Goldstini operators of dimension-8, \Eq{eq:S4chi}. We leave the analysis of these for section~\ref{sec:coll}.

\subsubsection*{Dipole Moments}

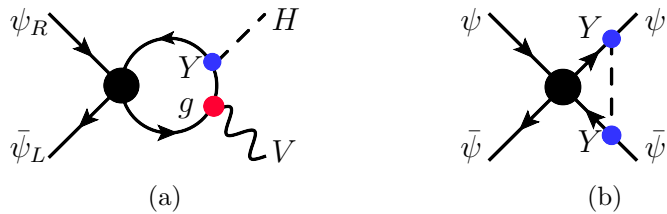
\begin{figure}[tp]
\centering
\hspace{1cm}\subfigure[]{\begin{picture}(100,80)
\put(0,0){ \includegraphics[height= 2cm]{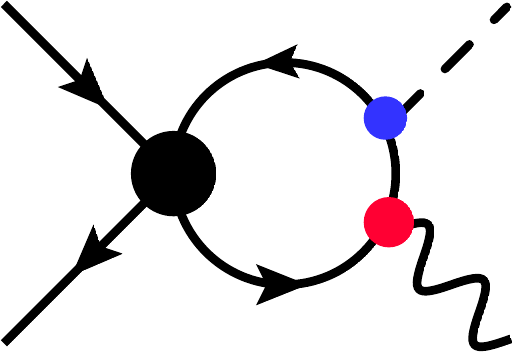}}
\put(-10,50){$\psi_R$}
\put(-10,2){$\bar\psi_L$}
\put(53,32){$Y$}
\put(53,19){$g$}
\put(88,50){$H$}
\put(88,0){$V$}
\end{picture}\label{fig2a}}
\hspace{2cm}
\subfigure[]{\begin{picture}(100,80)
\put(0,0){ \includegraphics[height= 2cm]{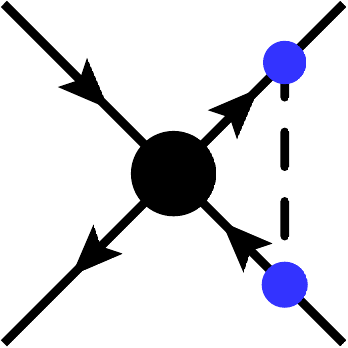}}
\put(-6,50){$\psi$}
\put(-6,2){$\bar\psi$}
\put(63,50){$\psi$}
\put(63,2){$\bar\psi$}
\put(38,47){$Y$}
\put(38,4){$Y$}
\end{picture}\label{fig2b}}
\caption{\footnotesize \emph{Some leading diagrams contributing to (magnetic) dipole moments and flavor transitions.}}
\label{fig:diags}
\end{figure}

When both $L$- \emph{and} $R$-handed SM fermions are pseudo-Goldstini, a Higgs and a gauge insertions are enough to saturate the necessary selection rules to generate dipole operators with $V = B, W, G$ (see e.g.~\Fig{fig2a}),  
\beq
\pmb{\psi}_L \textrm{ and } \pmb{\psi}_R \textrm{ Goldstini:} \quad c_{DV} \sim \frac{Y_\chi g_V}{m_*^2}\,.
\label{cDV}
\eeq
Instead, if only $\psi_L$ \emph{or} $\psi_R$ are composite, dipole operators can only be generated upon insertion of two extra Yukawa couplings and an elementary Higgs loop similar to the Barr-Zee one \cite{Barr:1990vd},
\beq
\pmb{\psi}_L \textrm{ or } \pmb{\psi}_R \textrm{ Goldstini:} \quad c_{DV} \sim \frac{Y_\chi^3 g_V}{(4\pi)^2 m_*^2}\,.
\label{cDVloop}
\eeq
The precise measurement of the anomalous magnetic moment of the muon \cite{Olive:2016xmw} sets a constraint on goldstino-compositeness of leptons
\beq
\pmb{\ell} \textrm{ and } \pmb{e}^c \textrm{ Goldstini:} \quad m_* \gtrsim 3.2 \TeV
\eeq
while if only $\pmb{\ell}$ or $\pmb{e}^c$ are pseudo-Goldstini the bound is negligible.

We note that our assumption on $R$-symmetry, which provides a concrete realization of the MFV paradigm, implies no new sources of CP violation beyond the SM. Therefore we obtain no relevant constraints from electric dipole moments.

\subsubsection*{Flavor transitions}

In MFV any flavor-violating operator is proportional to the SM Yukawa couplings.
This means that in the lepton sector any flavor transition is negligible due to the smallness of neutrino masses.
In the quark sector instead, MFV  implies that any flavor-violating process proceeds via the top, and the scale associated to certain operators needs to be above $1 \TeV$ \cite{Isidori:2012ts}.
Some of the potentially problematic operators, e.g.
\beq
\frac{e}{(4\pi)^2 m_*^2} \bar d_R Y_d^\dagger Y_u Y_u^\dagger \sigma^{\mu \nu} H^\dagger q_L F_{\mu \nu} \, , \quad \frac{g_*^2}{(4\pi)^4 m_*^2} (\bar q_L Y_u Y_u^\dagger \gamma_\mu q_L)^2 \,,
\label{mfv1}
\eeq
are generated at one or two weak (Higgs) loops respectively, so that the sensitivity of observables such as $B \to X_s \gamma$ or $\epsilon_K$ and $\Delta m_{B_d}$ is suppressed.
The strongest flavor constraints on goldstino-compositeness of quarks arise instead from the operator (see e.g.~\Fig{figb})
\beq
\frac{1}{m_*^2} (i H^\dagger {\lra { D_\mu}} H)( \bar q_L Y_u Y_u^\dagger \gamma^\mu q_L)
\label{mfv2}
\eeq
that contributes to $B \to X_s \ell^+ \ell^-$, $B_s \to \mu^+ \mu^-$, giving rise to the bound
\beq
\pmb{q} \textrm{ and } \pmb{u}^c \textrm{ Goldstini:} \quad m_* \gtrsim 2.3 \TeV \,,
\eeq
similar in size to the one from electroweak precision tests \Eq{qewpt}.

Comparable constraints arise if both $L$-handed quarks and leptons are pseudo-Goldstini. In such a case diagrams like \Fig{fig2b} generate the operators
\beq
\frac{g_*^2}{(4\pi)^2 m_*^2} (\bar q_L Y_u Y_u^\dagger \gamma_\mu q_L) (\bar \ell_L \gamma^\mu \ell_L) \, , \quad \frac{g_*^2}{(4\pi)^2 m_*^2} (\bar q_L Y_u Y_u^\dagger \gamma_\mu q_L) (\bar e_R \gamma^\mu e_R) \, ,
\label{mfv3}
\eeq
which also contribute to $B$-decays, implying
\beq
\pmb{q} \textrm{ and } (\pmb{\ell} \textrm{ or } \pmb{e}^c) \textrm{ Goldstini:} \quad m_* \gtrsim \left( 1.7 \textrm{ or } 2.7 \TeV \right) \times \left( \frac{g_*}{4 \pi} \right) \, .
\eeq

\subsubsection*{Other couplings}

Operators without external fermions can also be generated, simply because of the existence of a sector with a mass scale $m_*$ to which all SM fields  couple. Yet, such operators do not benefit from any strong coupling enhancement and provide, therefore, less relevant sensitivity to scenarios where (some) fermions are pseudo-Goldstini but the other species are elementary.

\subsubsection*{Other composites}

It is certainly plausible that, if SM fermions are indeed pseudo-Goldstini, the strong dynamics also involves in some way the Higgs boson (perhaps in a solution to the hierarchy problem), the transverse polarizations of vectors{\blu,} 
or perhaps it includes some of the fermions as non-Goldstini composites. These scenarios have been studied extensively already, in e.g.~Refs.~\cite{Giudice:2007fh,Liu:2016idz,Domenech:2012ai}, and their predictions in terms of EFTs are well known: composite fermions imply large $\Op_{4\psi}$, as already discussed in this work, while composite Higgs models imply large $\Op_H=(\partial^\mu |H|^2)^2$ and 
models {of composite gauge vectors} are characterized by large $\Op_{2W}=(D_\rho W_{\mu\nu}^a)^2$ and $\Op_{3W}=\epsilon_{abc}W^{a\, \nu}_{\mu}W^{b}_{\nu\rho}W^{c\, \rho\mu} $.

In this context, if the Higgs is composite the size of some of the Yukawa-induced operators discussed above is enhanced. 
In particular, operators that were generated via an elementary, thus weak, Higgs loop, see e.g.~\Fig{fig2b}, are no longer suppressed by $g_*^2/16\pi^2$, since for a composite Higgs the loop becomes strong.%
\footnote{This can also be understood by identifying, when the Higgs is composite, the Yukawa coupling in \Eq{expbreak} with a linear mixing \`a la \emph{partial compositeness}, see e.g.~Ref.~\cite{Liu:2016idz}.}
Such an enhancement applies to \Eqs{cDVloop} and (\ref{mfv1}), (\ref{mfv3}). Similarly, operators that were suppressed by $g_*^2/16\pi^2$ because either $\psi_L$ or $\psi_R$ were not pseudo-Goldstini, are enhanced as well when the Higgs is composite.

An interesting case is that in which the low-energy EFT includes non-SM light composite states, for which the selection rules differ. We have discussed in section~\ref{sec:emb} that embedding all of the quarks as Goldstini requires the existence of new color sextets $X$ and $Y$. 
Masses for these states require an extra source of explicit SUSY breaking, which we simply write as
\begin{equation}
m_{\textbf{6}_{2/3}} X_{-2/3}Y_{2/3}+   m_{\textbf{6}_{1/3}} X_{1/3} Y_{-1/3} \, .
\label{massXY}
\end{equation}
Being this a small deformation of the SUSY-preserving dynamics, we expect the sextets to be naturally light, thus present in the low-energy spectrum. 
Furthermore, the pseudo-Goldstini $X$ have SUSY-preserving interactions which, because of maximal $R$-symmetry, are invariant under $X\to-X$. Therefore such interactions do not contribute to the sextet's decay to the SM: symmetry breaking effects can play a major r\^ole here.
Specifically, a new SUSY-breaking spurion (with different quantum numbers than $g_{SM}$ or $m_{\mathbf{6}}$) can be introduced in association with a linear coupling of the sextet, the lowest-dimensional one being, schematically
\begin{equation}
\label{rhino}
\frac{\epsilon g_s}{m_*} \Psi \sigma^{\mu\nu} \chi G_{\mu\nu} \, ,
\end{equation} 
where $\Psi = Y_{2/3,-1/3}$ and $\chi = u^c, d^c$, where recall $Y$ is a non-Goldstini state. In fact, $Y$ could well be external to the strong dynamics, in which case its degree of compositeness should be factored in. Besides, $\epsilon \ll 1$ is expected since (\ref{rhino}) is an irrelevant operator. This operator is not only important for the sextet decay but also for its single production at the LHC, as we discuss in section \ref{sec:exoticXY}.

\section{Collider Phenomenology}
\label{sec:coll}

In section~\ref{sec:EFT} we have built the Goldstini EFT in the symmetric limit: Goldstini are characterized by interactions of dimension-8 that grow maximally fast with energy.
We have further established in section~\ref{sec:break} that the less irrelevant dimension-6 operators can only arise from (suppressed) SUSY-breaking effects. Constrains on the latter imply $m_*\gtrsim 2 \TeV$ for the compositeness scale in most of the interesting cases (with the exception of $R$-handed quarks, where the bounds are weaker).

In this section we attempt instead direct access to the strongly coupled dynamics by considering $2\to2$ pseudo-Goldstini scattering at the highest possible energies.
For comparison, we study both our scenario based on non-linearly realized SUSY and dimension-8 operators, as well as the standard one based on dimension-6 operators, which updates the analysis of Ref.~\cite{Domenech:2012ai}. Our main focus is LHC phenomenology, which gives us access to scattering of quarks in the multi-TeV region, but we also brush upon LEP-2 to investigate the extend to which leptons could arise as pseudo-Goldstini.

The results on composite quarks are summarized in Table~\ref{quarks}: we show the bounds on the SUSY-breaking scale $\sqrt{F}$ for goldstino-compositeness, as well as those on the scale $f$ for chiral-compositeness, in different composite quark scenarios.
A direct comparison between the two types of compositeness is also presented in terms of bounds on the strong coupling scale, identified as $m_* \equiv \sqrt{g_* F}$ for Goldstini and $m_* \equiv g_* f$ for chiral composites. 
The results on composite leptons, specifically $e_R$ and $\mu_R$, are given in \Eqs{boundseRb} and (\ref{boundsmubR}).

\subsection{LHC}
\label{sec:LHC}

\begin{figure}[t]
\begin{center}
\includegraphics[width=8cm]{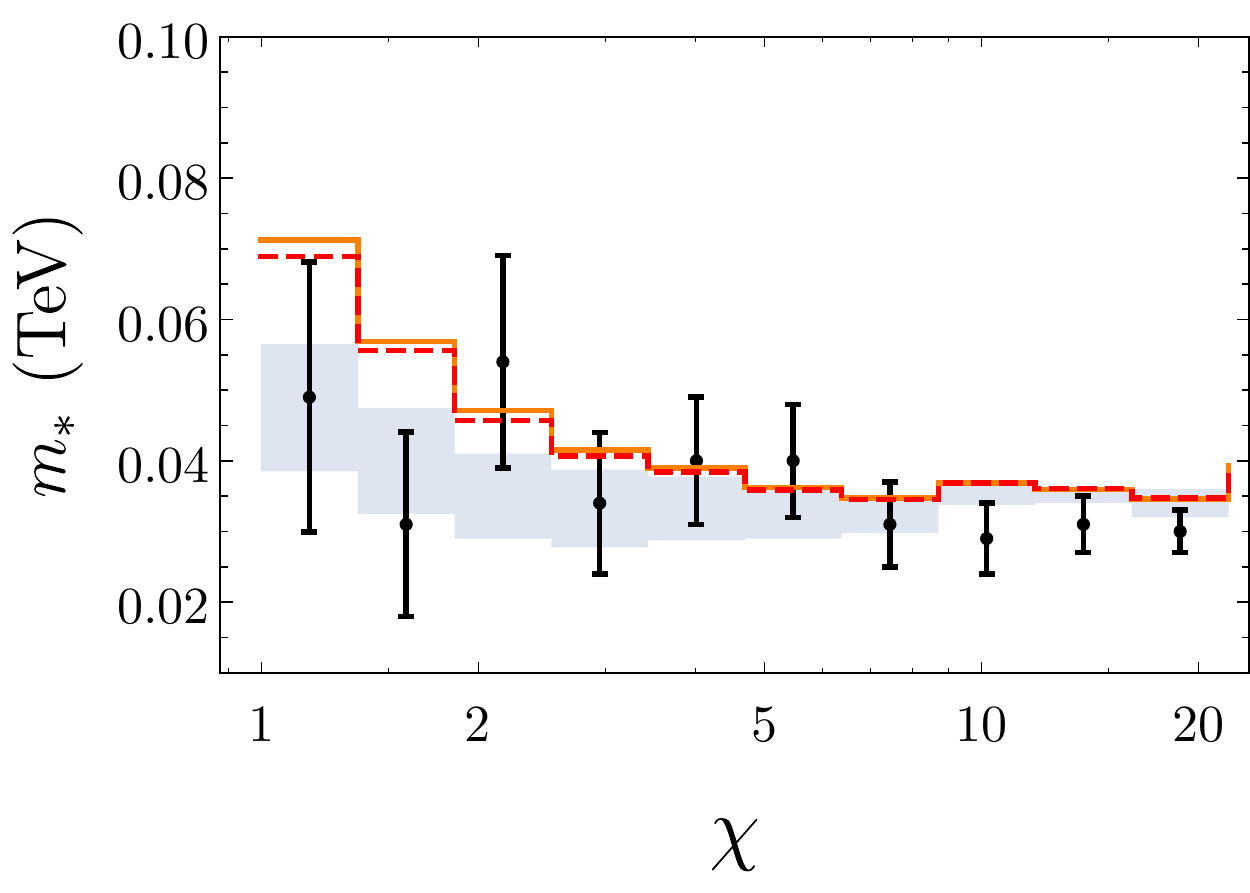}
\hspace{0.4cm}
\includegraphics[width=8cm]{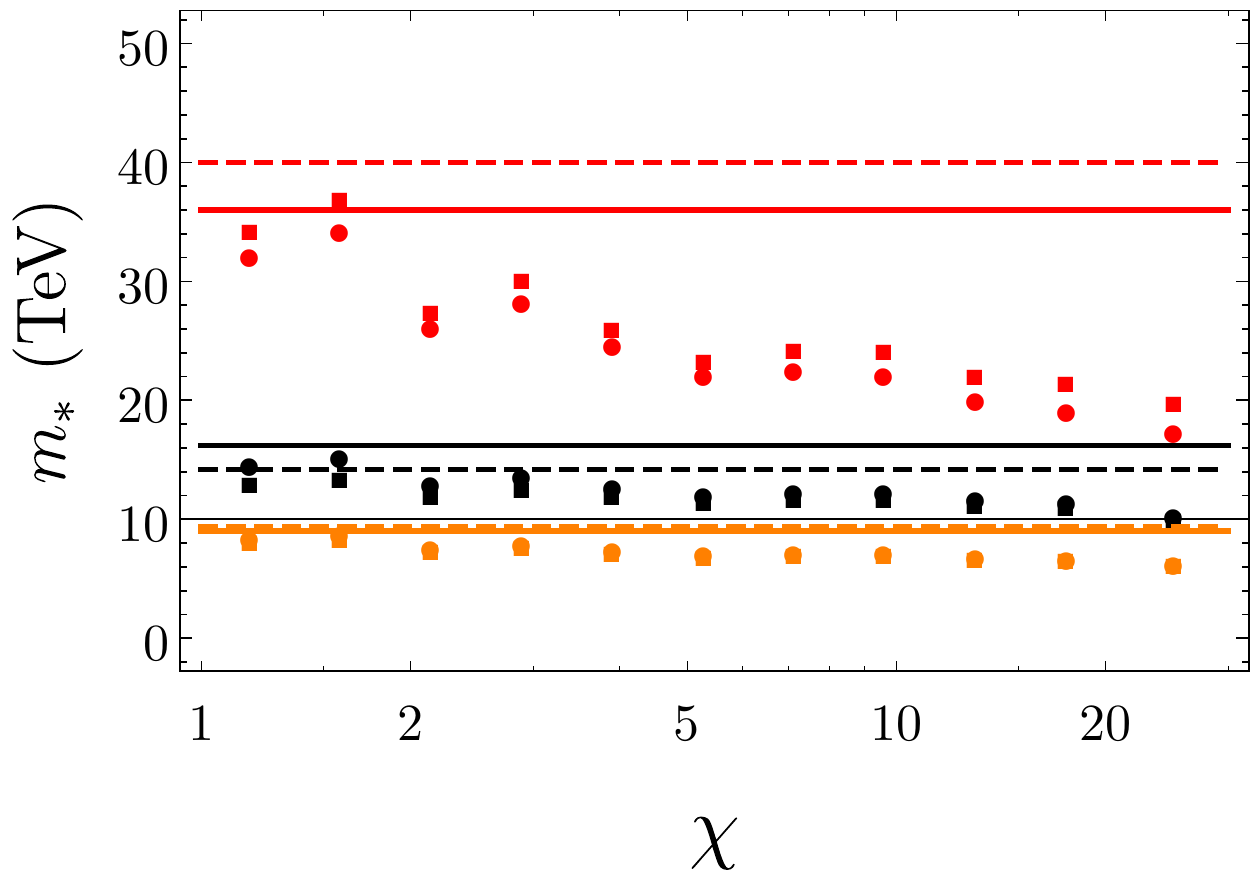}
\vspace{-1cm}
\end{center}
     \caption{\footnotesize\emph{Left: Dijet angular distributions in the highest-energy bin $M_{jj}>5.4 \TeV$, for the experimental data with its systematic plus statistical uncertainties (black points), the SM prediction with its theoretical error (blue band), and the BSM predictions for the dimension-8 operator $\mathcal{O}_{dd}$ (solid orange) or the dimension-6 operator $\mathcal{O}_{dd}^{(1)}$ \cite{Domenech:2012ai} (dashed red). The corresponding (positive) coefficients have been chosen to saturate the bounds \Eqs{bounds} and (\ref{boundsb}).
     Right: Bounds on the scale $m_*$ for a composite $d_R$ with a dimension-8 operator $|c_{dd}| = (4 \pi)^2/2m_*^4$ (orange) or a dimension-6 operator $|c_{dd}^{(1)}| = (4 \pi/m_*)^2$ (red) and for all quarks composite with dimension-8 operators $|c_{ij}| = (4 \pi)^2/2m_*^4$ (black), for different $\chi$ bins (dots) and for all bins combined (lines). The round dots and solid lines correspond to $c > 0$ while square dots and dashed lines to $c < 0$.}}
\label{fig:dist}\label{fig:bound}
\end{figure}

We focus on the process that gives, at present, access to the highest energies: dijet events $pp \to jj$ initiated by valence quarks at $13 \TeV$.
The experimental analysis can be found in Ref.~\cite{ATLAS:2016lvi}, corresponding to an integrated luminosity of $15.7 \,\mathrm{fb}^{-1}$ collected with the ATLAS detector.
The relevant pseudo-Goldstini interactions are parametrized by the operators
\begin{align}
&\Op_{uu} = (\partial_\nu \bar{u}_R \gamma^\mu u_R)(\bar{u}_R \gamma_\mu \partial^\nu u_R) &\quad\quad& 
\Op_{ud} = (\partial_\nu \bar{u}_R \gamma^\mu u_R)(\bar{d}_R \gamma_\mu \partial^\nu d_R) + \mathrm{h.c.} \nonumber\\
&\Op_{dd} = (\partial_\nu \bar{d}_R \gamma^\mu d_R)(\bar{d}_R \gamma_\mu \partial^\nu d_R) &\quad\quad&
\Op_{qu} = - (\partial_\nu \bar{q}_{L} \gamma^\mu q_{L}) (\partial^\nu \bar{u}_{R} \gamma_{\mu}  u_R) + \mathrm{h.c.} \nonumber\\
&\Op_{qq} = (\partial_\nu \bar{q}_L \gamma^\mu q_L)(\bar{q}_L \gamma_\mu \partial^\nu q_L) &\quad\quad& 
\Op_{qd} = - (\partial_\nu  \bar{q}_{L} \gamma^\mu q_{L}) (\partial^\nu \bar{d}_{R} \gamma_{\mu}  d_R) + \mathrm{h.c.} \, , 
\label{gops}
\end{align}
where in each particular composite quark scenario the Wilson coefficients are $c_{ij} = 1/(2F^2)$.
These operators contribute to the differential dijet cross section at the partonic level,%
\footnote{The leading contributions at high energies come from two initial first-family quarks (as opposed to quark-antiquark, two antiquarks or second- and third-family quarks), due to the enhanced PDFs in $pp$ collisions.}
\bea
\frac{d\sigma}{d \hat t}(q_iq_i\rightarrow q_iq_i)_{BSM} &=&
\frac{4\alpha_s}{9} \widehat A_1^{q_i} + \frac{1}{48\pi} \bigg[ \widehat B_1^{q_i}\, (2 \hat{u}^2 + 2 \hat{t}^2 + \hat{s}^2) + 6 \widehat B_2^{q_i} \, (\hat{u}^2 + \hat{t}^2) \bigg]\, ,\nonumber\\
\frac{d\sigma}{d \hat t}(ud\rightarrow ud)_{BSM} &=&
\frac{1}{16\pi} \hat{u}^2 \widehat B_3 \, ,
\label{bsmjj}
\eea
where $\hat s$, $\hat t$, $\hat u$ are the Mandelstam variables ($\hat s + \hat t + \hat u = 0$ in the massless approximation) and
\begin{equation}
\widehat A_1^{u,d} = c_{qq} + c_{uu,dd} \, , \quad
\widehat B_1^{u,d} = c_{qq}^2 + c_{uu, dd}^2 \, ,\quad 
\nonumber
\widehat B_2^{u,d} = c_{qu, qd}^2 \, , \quad
\widehat B_3 = c_{qq}^2 + c_{ud}^2 + c_{qu}^2 + c_{qd}^2 \, .
\end{equation}
The BSM contribution is then characterized by a strong energy growth and by being more central ($-t \approx s$) than the SM one, which instead peaks in the forward region because of $t$-channel gluon exchange. 
The analog of the CM scattering angle $\theta$ is conveniently represented by the boost-invariant difference between the two jet rapidities $y_{j}=\log \sqrt{(E_j+p_j^z)/(E_j-p_j^z)}$,
\begin{equation}
\chi\equiv e^{|y_{j_1}-y_{j_2}|}=(1+\cos\theta)/(1-\cos\theta) \, .
\end{equation}
The central region corresponds to small $\chi$, where the BSM differential cross section peaks, while the SM is approximately flat. Such a behavior can be recognized in \Fig{fig:dist} left panel, where we show the SM distribution and two different BSM contributions both corresponding to a composite $d_R$, one from goldstino-compositeness ($c_{dd} \neq 0$) and the other from chiral-compositeness ($c_{dd}^{(1)} \neq 0$, operators and cross section formulae can be found in Ref.~\cite{Domenech:2012ai}).

We compute the BSM particle-level cross section in two different ways to double-check: via the analytic differential cross section \Eq{bsmjj}, integrated over the PDFs (CT10 \cite{Lai:2010vv}) and binned according to the experimental analysis, and via a MadGraph \cite{Alwall:2014hca} simulation of our model implemented with FeynRules \cite{Alloul:2013bka}.
This is compared  with data from Ref.~\cite{ATLAS:2016lvi}, limited to highest-energy bin, with an invariant mass $M_{jj}>5.4 \TeV$; we check, a posteriori, that the bounds obtained on the compositeness scale $m_*$ are compatible with the EFT hypothesis: $m_*\gg M_{jj}$.
The analysis includes additional cuts on the transverse momentum of the leading jet $p_T(j_1)>450 \GeV$, on the average rapidity of the two jets $(y_{j_1}+y_{j_2})/2<1.1$, and on the rapidity difference $|y_{j_1}-y_{j_2}|/2<1.7$.
For the SM prediction we use the NLO differential distribution reported  in Ref.~\cite{ATLAS:2016lvi}, but we normalize it to the total number of SM events, compatible with the above cuts, that we compute using POWHEG \cite{Alioli:2010xa} and PYTHIA8 \cite{Sjostrand:2014zea} with PDF4LHC15 \cite{Butterworth:2015oua}, obtaining $\sigma_{SM}^{cuts}=50.8 \pm9.1 \,\mathrm{fb}$ (corresponding to $\sim 800$ events for the integrated luminosity of Ref.~\cite{ATLAS:2016lvi}).
NLO effects on the new physics distribution are instead neglected, although these could be relatively important compared to SM NLO effects, in the region of parameters that saturates the bounds (in this region, the SM and BSM contributions are similar in size, see \Fig{fig:dist}). 

Concerning the errors in our analysis, in the small $\chi$ region the dominant uncertainty is statistical, since the SM cross section is dominated by the forward region and therefore in the central region the statistics is small. Systematic uncertainties (dominantly from the jet energy scale) are relatively small. Theoretical uncertainties are sizable, yet smaller than the statistical at small $\chi$, becoming the dominant uncertainty at large $\chi$. Besides, even though the errors of the SM cross section are not gaussian, we treat them as such to simplify the analysis, symmetrizing the distributions such that the $1\sigma$ band remains unaltered.

We extract bounds on the Wilson coefficients in each $\chi$ bin and combine them to obtain the final constraint.
In \Fig{fig:bound} we report these bounds for the case of a composite $d_R$ as a function of $m_*$, defined as $c_{dd} \equiv \pm g_*^2/2m_*^4$ for goldstino-compositeness and $c_{dd}^{(1)} = \pm (g_*/m_*)^2$ for chiral-compositeness, with strong coupling $g_*=4\pi$. The final bounds for positive ($m_*^+$) or negative coefficients ($m_*^-$) at the $95\% \, \mathrm{CL}$ are%
\footnote{Goldstino operators are in fact subject to positivity constraints that require strictly positive coefficients (see section~\ref{sec:posit}). Our bounds for $c < 0$ simply illustrate the relevance of the interference with the SM.}
\begin{align}
\label{bounds}
&d_R \textrm{ chiral composite:} & & m_*^+ \gtrsim 36 \, (g_*/4\pi) \TeV \, , & & m_*^- \gtrsim 40 \, (g_*/4\pi) \TeV \,.
\\
&d_R \textrm{ Goldstini:} & & m_*^+ \gtrsim 9.4 \, \sqrt{g_*/4\pi} \TeV \, , & & m_*^- \gtrsim 9.0 \, \sqrt{g_*/4\pi} \TeV \,.
\label{boundsb}
\end{align}
Recall these bounds are derived from the bin with $M_{jj}>5.4 \TeV$. This means that for our analysis to be safely consistent with the EFT assumption $M_{jj} \ll m_*$, we must require $g_*\gtrsim 2$ in the case of chiral-compositeness and $g_*\gtrsim 4.5$ for goldstino-compositeness, i.e.~the new physics must be strongly coupled.

\begin{table}[t]
\begin{center}
\begin{tabular}{ccc}
\hline
 & goldstino & chiral \\
composites & \qquad $\,\,\,\sqrt{F}$ $\,\,[m_*]$ ($\!\TeV$) & \qquad \quad $\,\,f$ $\,\,[m_*]$ ($\!\TeV$) \\
\vspace{-0.5cm} \\
\hline \\
\vspace{-1cm} \\
$d_R$ & $ 2.6 $ $ [9.4] $ & $ 2.9 $ $ [36] $  \\
$u_R$ & $ 3.8 $ $ [13.5] $ & $ 4.7 $ $ [59]$  \\
$u_R, d_R$ & $ 3.9 $ $ [13.7] $ & $ 4.9 $ $ [62] $  \\
$q_L$ & $ 3.9 $ $ [13.7] $ & $ 4.9 $ $ [62] $\\
$q_L, d_R$ & $ 4.0 $ $ [14.2] $ & $ 5.0 $ $ [63] $\\
$q_L, u_R$ & $ 4.5 $ $ [16.1] $ & $ 5.7 $ $ [72] $\\
$q_L, u_R, d_R$ & $ 4.6 $ $ [16.2] $ & $ 5.8 $ $ [73] $  \\
\hline
\end{tabular} \\
\caption{\footnotesize\emph{$95\% \, \mathrm{CL}$ bounds on the SUSY-breaking scale $\sqrt{F}$ for different quark as pseudo-Goldstini scenarios (second column) and on the scale $f$ for different cases of chiral-compositeness of quarks (third column). The respective Wilson coefficients are given by $c_{ij} = 1/(2F^2)$ for the operators in \Eq{gops} and by $c_{ij}^{(1)} = 1/f^2$ for the operators $\mathcal{O}_{ij}^{(1)} = (\bar \psi_i \gamma^\mu \psi_i) (\bar \psi_j \gamma_\mu \psi_j)$ given in Ref.~\cite{Domenech:2012ai}. The compositeness scale in each case is given by $m_* \equiv \sqrt{g_* F}$ or $m_* \equiv g_* f$ with $g_* = 4 \pi$, their bounds reported in brackets.}}
\label{quarks}
\end{center}
\end{table}

Besides, we discussed in section~\ref{sec:break} the existence of dimension-6 four-fermion operators $\Op_{4\psi}$ generated from explicit SUSY-breaking. The question arises then whether experimental searches, in the case at hand $2\to2$ quark scattering at the LHC, are more sensitive to these more relevant (but $g_{SM}$-suppressed) effects or to the dimension-8 (but $g_*$-enhanced) effects. \Eqs{bounds} and (\ref{boundsb}) show that, for $g_*$ within the validity of the EFT, the constraints on pseudo-Goldstini from the latter are clearly superior.
As a matter of fact, bounds on goldstino-compositeness from the LHC are stronger than any indirect ones from SUSY-breaking effects. This is clearly due to the large energies accesible at this collider (roughly $E/m_* \gg g_{SM}/g_*$).

Finally, the scenario of a composite $d_R$ is constrained the weakest of all the cases, while when all quarks are composite, the LHC reach is maximal. In \Fig{fig:bound} we show the constraints in this scenario, the final result being
\begin{align}
\label{bounds2}
&q_L, u_R, d_R \textrm{ Goldstini:} & & m_*^+ \gtrsim 16.2 \, \sqrt{g_*/4\pi} \TeV \, , & & m_*^- \gtrsim 14.2 \, \sqrt{g_*/4\pi} \TeV \,.
\end{align}
The bounds for the rest of composite scenarios are given in Table~\ref{quarks}.

\subsection{LEP}
\label{sec:LEP}

During its second phase, LEP collided electrons and positrons at energies significantly higher than the $Z$-pole, measuring angular distributions with percent precision.
Here we focus on how the differential cross sections for $e^+e^-\to e^+e^-, \mu^+ \mu^-$ are affected by the goldstino-compositeness of the $R$-handed electron $e_R$ and muon $\mu_R$, and compare them with LEP-2 data \cite{Schael:2013ita} to extract bounds on their SUSY-breaking scale $F$ or compositeness scale $m_*$.

Let us first recall that limits on standard chiral-compositeness of $e_R$, parametrized by the dimension-6 operator $c^{(6)}(\bar e_R \gamma^\mu e_R) (\bar e_R \gamma_\mu e_R)$, were extracted by the LEP collaborations,
\begin{align}
&e_R \textrm{ chiral composite:} & &  m_*^+ \gtrsim 43 \, (g_*/4\pi) \TeV \, , & & m_*^- \gtrsim 40 \, (g_*/4\pi) \TeV \,,
\label{boundseR}
\end{align}
where we normalized the corresponding Wilson coefficient as $|c^{(6)}| = (g_*/m_*)^2$.
Bounds were obtained as well on $(\bar e_R \gamma^\mu e_R) (\bar \mu_R \gamma_\mu \mu_R)$: $m_*^+ \gtrsim 41 \, (g_*/4\pi) \TeV$ and $m_*^- \gtrsim 33 \, (g_*/4\pi) \TeV$.

Our interest here is in deriving first-time bounds on the goldstino-compositeness of $e_R$, which is parametrized by the dimension-8 operator
\begin{equation}
\mathcal{O}_{ee} = (\partial_\nu \bar{e}_R \gamma^\mu e_R)(\bar{e}_R \gamma_\mu \partial^\nu e_R) \,,
\label{geops}
\end{equation}
and on the scenario where both $e_R$ and $\mu_R$ are pseudo-Goldstini, in which besides $\mathcal{O}_{ee}$ also
\begin{equation}
\mathcal{O}_{e\mu} = (\partial_\nu \bar{e}_R \gamma^\mu e_R)(\bar{\mu}_R \gamma_\mu \partial^\nu \mu_R) + \mathrm{h.c.}
\label{gmuops}
\end{equation}
is generated, with the same coefficient.%
\footnote{We do not study $\tau_R$ compositeness since the sensitivity of LEP in $e^+e^-\to \tau^+ \tau^-$ is weaker.}
These operators induce non-standard contributions to the differential cross sections for Bhabha scattering and dimuon production -- their analytic expressions are reported in the Appendix.
They share some similarities with the dijet case, in particular the $t$-channel photon exchange also gives rise to a forward singularity, such that the SM contribution and the interference with the BSM are enhanced for small angles.

The experimental sensitivity in $e^+ e^-$ production at small angles is approximately $4\%$ ($95\% \, \mathrm{CL}$) of the SM contribution. This means that, contrary to the LHC, LEP is really testing small departures from the SM, and it is sensitive then to the SM-BSM interference term.
We take the SM prediction provided in Ref.~\cite{Schael:2013ita} and compute the new physics effects analytically, see the expressions in the Appendix.
The theoretical uncertainties on the total SM cross section amount to $2\%$ for $\sigma(e^+ e^-)$ and $0.4\%$ for $\sigma(\mu^+ \mu^-)$, resulting in an uncertainty for $d\sigma/d\cos\theta|_{SM}$ of $4\%$ and $1\%$ respectively (assuming the error to be uniformly distributed with the scattering angle $\theta$).
For the purpose of this analysis, we use samples of events with integrated luminosity of $3\,\mathrm{fb}^{-1}$ and increasing effective CM energy from 189 to 207 GeV.%
\footnote{Initial-state photon radiation may reduce the CM energy of the dilepton production. In the LEP analyses only events with soft initial-state radiation are retained \cite{Schael:2013ita}.} 

We combine the limits from different energy and angular bins and obtain
\begin{align}
\label{boundseRb}
&e_R \textrm{ Goldstini:} & & m_*^+ \gtrsim 1.8 \, \sqrt{g_*/4\pi} \TeV \, , & & m_*^- \gtrsim 1.4 \, \sqrt{g_*/4\pi} \TeV \,.
\\
&e_R, \mu_R \textrm{ Goldstini:} & & m_*^+ \gtrsim 1.9 \, \sqrt{g_*/4\pi} \TeV \, , & & m_*^- \gtrsim 1.5 \, \sqrt{g_*/4\pi} \TeV \,.
\label{boundsmubR}
\end{align}
The bounds on the scenario where both electron and muon are pseudo-Goldstini is driven by Bhabha scattering, with a SM cross section significantly larger than dimuons -- the constrains on only $\mathcal{O}_{e\mu}$ are milder $m_*^+ \gtrsim 1.6 \, \sqrt{g_*/4\pi} \TeV$ and $m_*^- \gtrsim 1.5 \, \sqrt{g_*/4\pi} \TeV$.
It is amusing to find that goldstino-compositeness of leptons is allowed at incredibly small scales from direct searches, even for light leptons and maximally strong coupling $g_* = 4\pi$.

In fact, precision is relatively less important, compared to the collider energy, when searching for this type of compositeness, as it becomes clear when comparing the reach of LHC vs LEP. 
In contrast, dimension-6 operators are typically better constrained by LEP (or barely so by the LHC \cite{Farina:2016rws}).
Indeed, the relative size of our effects scales as $\delta \sim (E/m_*)^4$ in the linear regime, and as $(E/m_*)^8$ for large deviations from the SM. In order to increase the bound on the scale $m_*$ by a factor of 2 we would need to increase the precision at a given energy by at least a factor of 16; the same goal can be achieved by a factor of 2 increase in energy. The LHC high-energy reach makes it the best machine to test for fermions with enhanced soft behavior. 
Another consequences of this same fact is that the limits on the scale $m_*$ for chiral-compositeness are much higher, their effects scaling as $(E/m_*)^2$.
This also means that, for a proper extraction of the bounds, SUSY-breaking effects which generate four-fermion operators should be included in the analysis \cite{Falkowski:2015krw}, even if suppressed by $(g_{SM}/g_*)^2$.

\subsection{Positivity Constraints}
\label{sec:posit}

An interesting aspect of the dimension-8 interactions studied in this work is that they are subject to positivity constraints. Indeed, the basic requirement of unitarity in the underlying theory, together with analyticity of the $2\to2$ scattering amplitudes, implies that the Wilson coefficients of the operators in \Eqs{gops} and (\ref{geops}), (\ref{gmuops}), be strictly positive \cite{Adams:2006sv,Bellazzini:2016xrt}.

From a phenomenological perspective, this represents an important prior, from first principles, to our statistical analysis, that reduces the parameter space by half.
Without any prior, our analysis above leads to $95\% \, \mathrm{CL}$ intervals of the form $[-c^-(g_*,m_*^-),c^+(g_*,m_*^+)]$, while our theory prior implies $]0,\hat c(g_*,m_*)]$. 
Taking it into account in our statistical analysis, we find,
\begin{align}
&d_R \textrm{ Goldstini:} & & m_* \gtrsim 9.3 \, \sqrt{g_*/4\pi} \TeV \, , \label{boundspos} \\
&e_R \textrm{ Goldstini:} & & m_* \gtrsim 1.7 \, \sqrt{g_*/4\pi} \TeV \, . \label{boundseRpos}
\end{align}
Note that these limits do not improve our knowledge on goldstino-compositeness (compared to \Eqs{boundsb} and (\ref{boundseRb})). On the contrary, these more conservative bounds highlight the importance of keeping the prior, not to overexclude the physically consistent region of parameter space.

Besides, whether $\hat c(g_*,m_*)<c^{+}(g_*,m_*^+)$ depends on the likelyhood $L$ being symmetric or not under reflection of the Wilson coefficients $c\to -c$.
Indeed, if it is symmetric then
\begin{equation}
0.95=\frac{\int_{c^-}^{c^+}dc \, L}{\int_{-\infty}^{\infty}dc \, L}=\frac{\int_{0}^{c^+}dc \, L}{\int_{0}^{\infty}dc \, L} \, ,
\end{equation}
where the last expression defines the bound with our prior, thus $\hat c = c^{+}$.
Under our assumption of symmetric errors, a significant asymmetric likelyhood can arise if two conditions are met: an under or upper fluctuation in the data is present and the size of the constraints is such that the cross section has a sizable interference term (linear in the Wilson coefficient).
In our LHC analysis, departures from the SM are not large (apart from the third $\chi$ bin in \Fig{fig:dist}). Most importantly, the experimental resolution, limited by statistical errors at present, is not enough to resolve the SM-BSM interference term in the cross section, which is instead dominated by the quadratic new physics contribution.%
\footnote{Notice that this fact is safely compatible with the validity of our EFT expansion, due to the underlying strong coupling, see the discussion below \Eq{boundsb} and Ref.\cite{Contino:2016jqw}.}
For this reason, positivity constraints do not improve sizably our constraints on $m_*$ and $g_*$.

\subsection{Outlook - Dibosons}
\label{sec:diboson}

In the previous sections we have shown that goldstino-compositeness is described by dimension-8 operators, and that currently the LHC is more sensitive to such strongly coupled effects than to SUSY-breaking effects, even though the latter give rise to lowest-dimensional interactions.
 
If other species in the SM are involved in the strong dynamics, there can also be large effects in other LHC processes, beyond $2\to 2$ fermion scattering, that are however unique to pseudo-Goldstini.
Interestingly, these effect are characterized by dimension-8 operators as well, as shown in Tables \ref{tab1:indepcoup} and \ref{tab2:modedepcoup}. 
Indeed, if the Higgs is composite in addition to the light quarks, the operator
\beq
\frac{1}{2F^2}\left(i\bar \psi \gamma^{\{\mu} \partial^{\nu\}} \psi+\mathrm{h.c.}\right)\partial_\mu H^{\dagger} \partial_\nu H
\eeq
modifies the amplitudes for $h$ pair production, as well as $W_L^+W_L^-$ and $Z_LZ_L$ ($L=$ longitudinal). Similarly, if the transverse (T) polarizations of vectors are strongly interacting, along the lines of 
{Ref.}~\cite{Liu:2016idz}, the amplitudes for $W_T^+W_T^-$, $Z_TZ_T$, $Z_T\gamma$ and $\gamma\gamma$ are modified by
\beq
-\frac{1}{4F^2}F^A_{\mu\nu}F^{A\,\mu}_{\phantom{\mu}\rho}\left(i \bar \psi \gamma^{\{\rho}\partial^{\nu\}} \psi+\mathrm{h.c.}\right) \,.
\eeq
This is very interesting because, even in a completely model-independent approach, the amplitudes for processes with neutral gauge boson final states are not modified at the level of dimension-6 operators; experimental constraints from $ZZ$ final states are at present already derived in terms of dimension-8 operators of the form $iH^\dagger \overset{\text{\scriptsize$\leftrightarrow$}}{D}_{\mu} HD^\nu B_{\nu\rho}B^{\mu\rho}$.
This kind of operators are typically subleading in theories without symmetries and therefore these type of searches have received so far little attention.
Pseudo-Goldstini offer instead a context where all dimension-6 operators are naturally suppressed by symmetries and dimension-8 effects are naturally  leading, so that these searches  could play the most important r\^ole.

Another reason why the  operators we propose in this article are interesting is the following.
Currently the entire new physics parametrization of neutral diboson final states, see Ref.~\cite{Gounaris:1999kf},  only induces final states with one longitudinal and one transverse vector.%
\footnote{This is due to Ref.~\cite{Gounaris:1999kf} providing a parametrization for anomalous neutral triple gauge couplings that limit the process to take place through the $s$-wave, which forbids in this case identical bosons in the final state; in our case higher waves are allowed and different final states open.} 
The corresponding amplitudes decrease at high energy as $1/E$ compared to the amplitudes for $TT$ or $LL$ final states, and is therefore typically subdominant. Instead, the effects we advocate here modify all $TT$ and $LL$ amplitudes (including the one with $(+,-)$ helicity that dominates in the SM) which, beside being more generic, will also be easier to find.

In summary, SM fermions as pseudo-Goldstini provide the first structurally motivated scenario where processes with neutral gauge boson pair production can be used as valuable BSM search tools. Importantly their parametrization departs from that traditionally adopted in these searches, and involves a richer variety of phenomena with enhanced high-energy behavior and larger SM-BSM interference. We leave this for future work.

\section{Phenomenology of the new colored states}
\label{sec:exoticXY}

In section~\ref{sec:EFT}, we have found that models with all quarks as pseudo-Goldstini and maximal $R$-symmetry require the existence of new exotic colored particles and their extra (non-SM fermions and not pseudo-Goldstini) partners. Their quantum numbers under the gauge and flavor symmetries $U(1)_Y\times SU(2)_{L}\times SU(3)_C \times SU(3)^{Flav}_{u} \times SU(3)^{Flav}_{d}$ are
\begin{equation}
\pmb{X}_{-{2\over3}}=(\mathbf{1},\mathbf{6},\mathbf{3},\mathbf{1})_{-{2\over3}}\,,\quad \pmb{X}_{{1\over3}}=(\mathbf{1},\mathbf{6},\mathbf{1},\mathbf{3})_{{1\over3}}\,,\quad \pmb{Y}_{{2\over3}}=(\mathbf{1},\mathbf{6}^*,\mathbf{3}^*,\mathbf{1})_{{2\over3}}\,,\quad \pmb{Y}_{-{1\over3}}=(\mathbf{1},\mathbf{6}^*,\mathbf{1},\mathbf{3}^*)_{-{1\over3}}
\end{equation}
They have the same quantum numbers as $u^c$ and $d^c$ except they transform as a $\mathbf{6}$ of $SU(3)_C$. 
Their Dirac masses $m_{\textbf{6}_{2/3,1/3}}$ are naturally small, as they break SUSY explicitly, see \Eq{massXY}.
In the following we will always omit the hypercharge of $X$ and $Y$, being always understood its value from the coupling to the SM quarks.

\subsubsection*{Decays}

These sextets can couple to gluons and the SM $u$ or $d$ quarks through the model-dependent SUSY-breaking couplings in \eq{rhino} (recalling that $\mathbf{6}\otimes \mathbf{3}=\mathbf{8}\oplus\mathbf{10}$ and $\mathbf{3}\otimes\mathbf{8}=\mathbf{3}\oplus\mathbf{6}^*\oplus\mathbf{15}$)
\begin{align}
\left(\frac{\epsilon g_s}{m_*}\right)\,\overline{K}^{\phantom{i}b\,A}_{i} Y^i \sigma^{\mu\nu} q_b G_{A\, \mu\nu} + \mathrm{h.c.} \,, \quad q = u^c, d^c
\label{opYq}
\end{align}
where $\overline{K}^{\phantom{i}b\,A}_{i}$ are the Clebsch-Gordan coefficients linking the anti-sextet to the anti-triplet and the adjoint, normalized such that
\begin{equation}
\label{unitarity}
\overline{K}^{\phantom{i}b\,A}_{i}K_{\phantom{i}A\,b}^{j} = \delta_i^{\phantom{i}j} \,, \qquad \overline{K}_i^{\phantom{i}b\,A} K^i_{\phantom{i}A\,b} = 6 \,.
\end{equation}
$\epsilon$ is a SUSY-breaking parameter that depends on assumptions like the degree of compositeness of $Y$. 
Despite the interactions (\ref{opYq}) being suppressed, they represent the main decay mode for the sextet.
In particular, they open the following decay channels:
\begin{equation}
X, Y \rightarrow jj, \, bj, \, tj
\label{decaychannels}
\end{equation}
with decays widths approximately given by,
\begin{equation}
\Gamma(Y\rightarrow q g) \approx \alpha_s\epsilon^2 \frac{m_\textbf{6}^3}{4m_*^2}\,.
\label{widthsXY}
\end{equation}
We assume, consistently with $\epsilon$ being a small symmetry-breaking parameter, that $\Gamma \ll m_{\textbf{6}}$, such that the narrow-width approximation applies to the searches described below.
  
\subsubsection*{Production and direct searches}

The sextet can either be singly produced via $gq \rightarrow Y$ (see e.g.~Ref.~\cite{Han:2010rf}) or pair produced 
$gg, q \bar q \rightarrow X\bar{X}, \, Y \bar{Y}$.
We assume $m_{\textbf{6}_{2/3}} = m_{\textbf{6}_{1/3}}$ but focus on direct searches of a single sextet flavor, coupled to first generation quarks (for a fully degenerate family spectrum, our analysis can be appropriately rescaled).
The relative sensitivity of single production is larger for heavy sextets, although it requires large couplings $\epsilon/m_*$, while double production presents a poorer mass reach, but it is more model-independent given the cross section is fixed by QCD.

\begin{figure}[t]
\begin{center}
\includegraphics[height=7.0cm,valign=t]{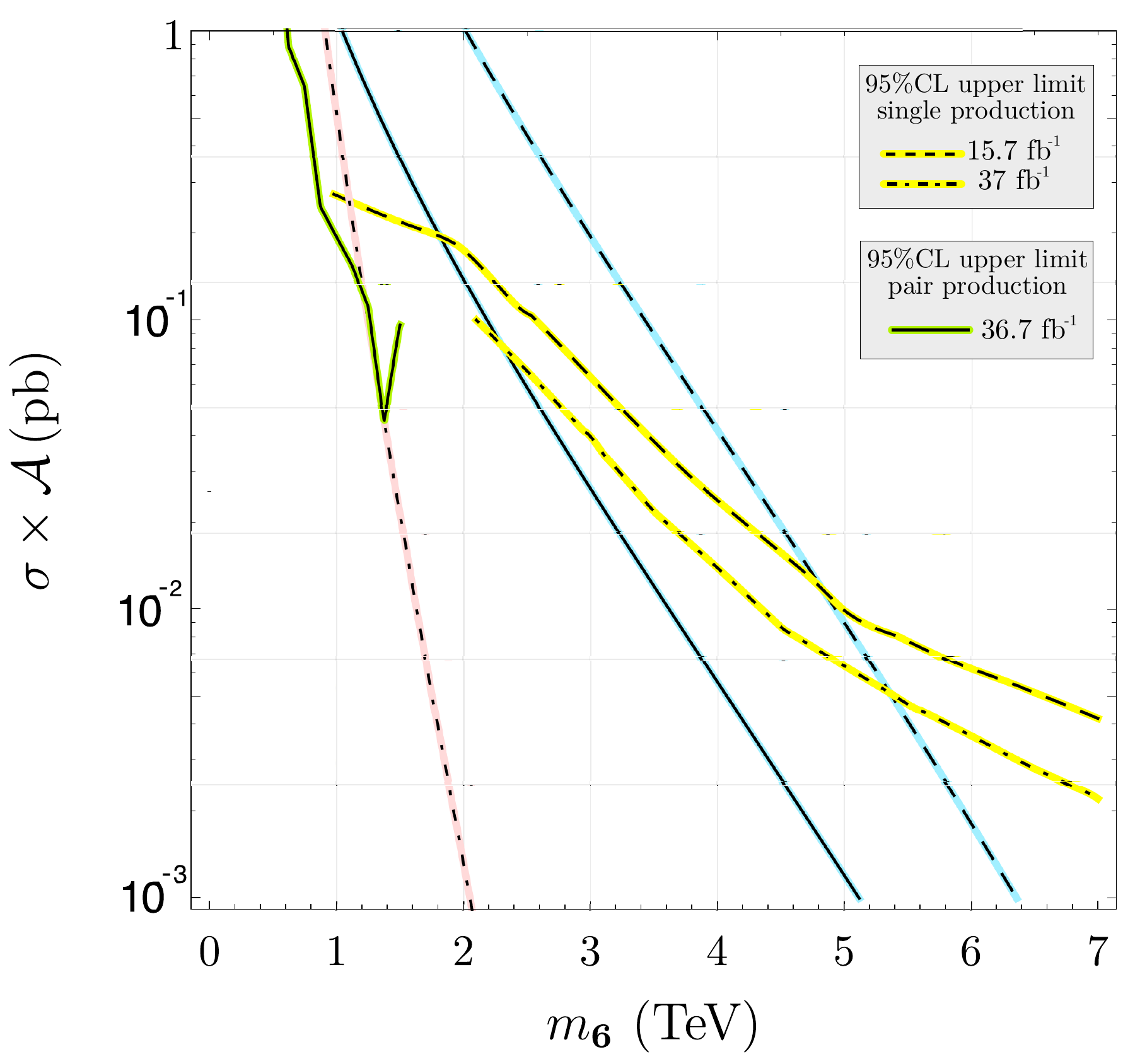}
\hspace{1cm}
\includegraphics[height=7.5cm,valign=t]{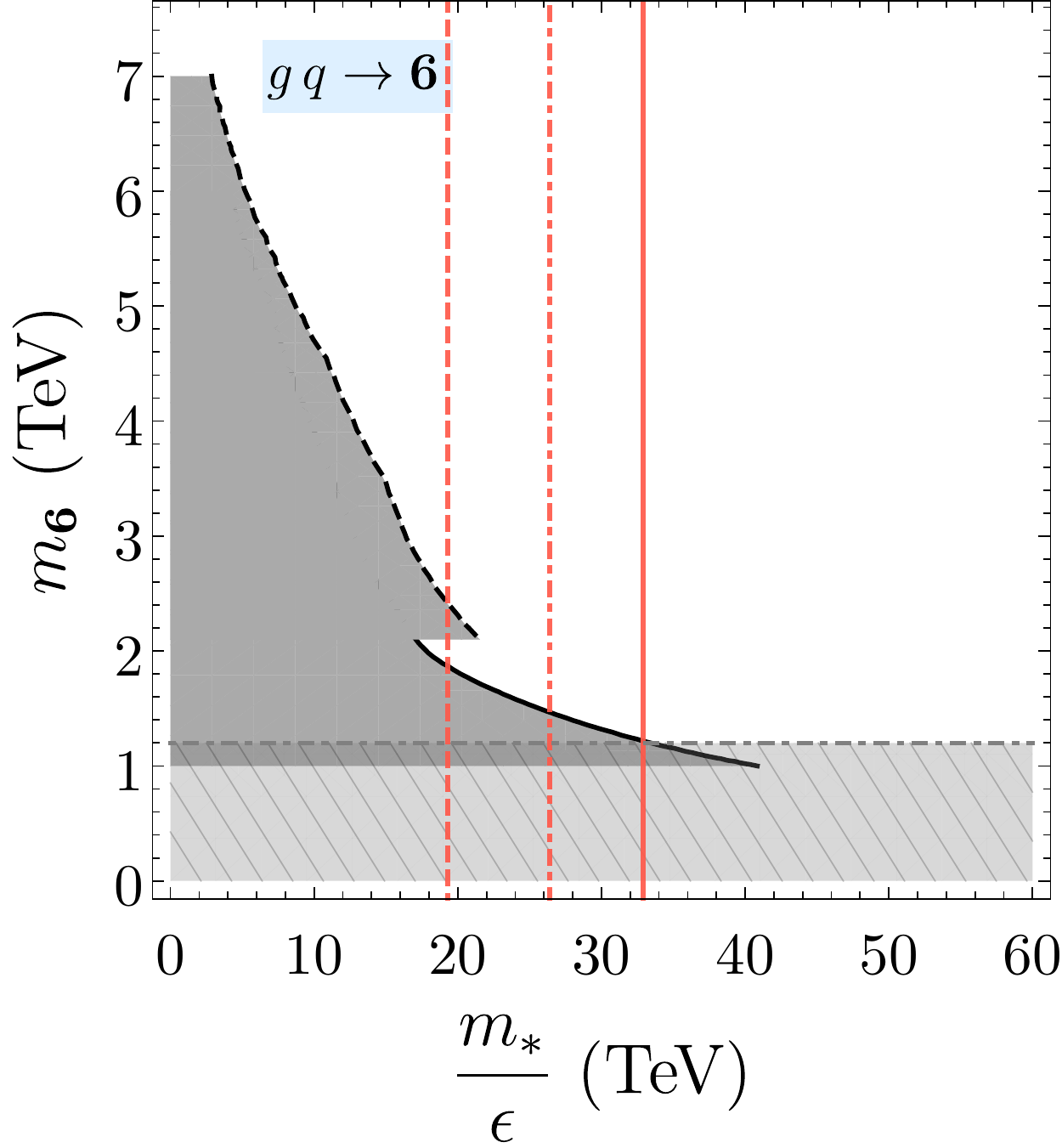}
\vspace{-0.3cm}
\end{center}
\caption{\footnotesize\emph{Left: $95\% \, \mathrm{CL}$ upper limits on single- and pair-production cross sections, compared with the theoretical cross sections for single-production (blue) with $m_*/\epsilon = 20$ (solid) or $m_*/\epsilon = 10$ (dashed) and pair-production (dot-dashed red).
Right: Exclusion regions at $95\% \, \mathrm{CL}$ from single-production (dark grey; \cite{ATLAS:2016lvi} solid line and \cite{Aaboud:2017yvp} dashed line) and double-production (light grey; dot-dashed line). Also shown are lower bounds (red) on the (inverse) coupling of the sextet, $m_*/\epsilon \approx 19 \TeV$ (dashed; corresponding to a scale $m_* = 5.4 \TeV$, the highest energy of our dijet analysis, and a small parameter $\epsilon = 0.3$), $m_*/\epsilon \approx 27 \TeV$ (dot-dashed; same $m_*$, $\epsilon = 0.2$) and $m_*/\epsilon \approx 33 \TeV$ (solid; coupling below which pair production dominates the constraints).}}
\label{fig:bound6}
\end{figure} 

The LO partonic cross section for single production and decay is given by
\begin{equation}
\sigma_{gq \rightarrow Y \rightarrow jj } = \sigma_{gq \rightarrow Y} \cdot \text{BR}(Y\rightarrow jj) \, , \quad
\sigma_{gq\rightarrow Y} = \frac{\pi^2\alpha_s}{2}\left(\frac{\epsilon}{m_*}\right)^2m_\textbf{6}^2 \, \delta(\hat{s}-m_\textbf{6}^2).
\label{xsecsixplet}
\end{equation}
where $q = u, d$. This cross section suffers from significant NLO corrections (e.g.~from $gg \rightarrow Y j$) that we neglect given the exploratory nature of our analysis.
We compare our LO prediction, appropriately convoluted with the PDFs, with experimental searches of singly produced exotic quark-like resonances decaying into dijets \cite{ATLAS:2016lvi,Aaboud:2017yvp}.

Pair production gives rise to a four-jet signal. We compute the associated cross section at LO with MadGraph \cite{Alwall:2014hca}, requiring four leading jets with $p_T > 80 \GeV$ and pseudo-rapidity $|\eta|<1.4$. We compare our results with the cross section bounds provided by Ref.~\cite{ATLAS:2017gsy}.

We present in \Fig{fig:bound6} left panel single- and pair-production cross sections along with their corresponding experimental limits. For single production we take two different values of the (inverse) coupling $m_*/\epsilon = 10, \, 20 \TeV$ in order to illustrate the variation of the cross section.
In the right panel we show the final bounds in the $(m_*/\epsilon, m_\textbf{6})$ plane. As anticipated, single production dominates the constraints at high masses and large couplings (i.e.~small $m_*/\epsilon$). 
Note that given the sextet is predicted upon all quarks being pseudo-Goldstini (and maximal $R$-symmetry), for which LHC searches in section~\ref{sec:LHC} set a bound on the compositeness scale $m_* \gtrsim 16.2 \sqrt{g_*/4\pi} \TeV$ (see Table~\ref{quarks}), the coupling $\epsilon/m_*$ cannot be too large while keeping $\epsilon \ll 1$. This is exemplified by the vertical (red) lines in \Fig{fig:bound6}, which should be understood as upper bounds on the sextet linear coupling, restricting the regime where limits from single production apply.
In contrast, pair production sets the robust lower bound $m_\textbf{6} \gtrsim 1.2 \TeV$.

\section{Conclusions and Outlook}

The quest for an answer to the question \emph{``what is matter made of?''} has always been at the heart of particle physics research. 
To keep pursuing this endeavor, the priority of the future high-energy experimental program, both at the High-Luminosity LHC and future colliders, is to understand to what extent those that we call SM particles are indeed elementary entities, or to find signs of their substructure.
\vspace{2mm}

Unitarity of the underlying theory implies there are only two {relativistic effective field theories} 
for fermion compositeness, which can be broadly differentiated from the point of view of a long-distance observer. The first such scenario is characterized by a series of effective four-fermion operators of dimension-6, whose phenomenological implications are well known and studied in the literature. These operators preserve chiral symmetry, providing a structural reason why the new strong interaction associated with compositeness does not percolate to the SM fermion masses and Yukawas.
The second scenario has instead a well-defined limit where the dimension-6 effects vanish and compositeness is manifested through dimension-8 operators. One certain realization is associated with non-linearly realized supersymmetry; that is, the fermions are composite pseudo-Goldstini.

In this article we built the EFT of ${\cal N}$ Goldstini and discussed how to embed the SM fermions in such a framework. In particular, the SM gauge and flavor groups are subgroups of the $R$-symmetry $\subset U({\cal N})$. The SM gauge and Yukawa couplings explicitly break supersymmetry, these breakings propagating to other observables -- in particular they generate the dimension-6 operators that were forbidden in the exact supersymmetric limit -- but can be treated as perturbative and we estimated their sizes. 

We compared this goldstino-compositeness with the standard chiral-compositeness, confronting them both with experimental data in the form of measurements of dijet distributions at the LHC or $e^+e^-$ scattering at LEP (in addition to other low-energy observables). Wilson coefficients of the standard four-fermion operators scale as $\big(g^{(6)}_*/m^{(6)}_*\big)^2$, while the dimension-8 as $(g^{(8)}_*)^2/(m^{(8)}_*)^4$, in terms of the physical scales and couplings of the new dynamics. 
Then, an experiment performing at energy $E_{exp}$ provides constraints that naively relate as,
\beq
m^{(8)}_* \sim m^{(6)}_* \left(\frac{E_{exp}}{m^{(6)}_*}\right)^{1/2} \left(\frac{g^{(8)}_*}{g^{(6)}_*}\right)^{1/2} \,,
\eeq
an intuition that we confirmed with dedicated analyses. Since ${E_{exp}}/{m^{(6)}_*} \ll 1$ in a sensible EFT, for $g^{(8)}_* \sim g^{(6)}_*$ bounds on goldstino-compositeness are always poorer than those on chiral-compositeness: by a factor of 4 for light quarks (tested at LHC) and by a factor of 20 for light leptons (tested at LEP-2). For pseudo-Goldstini, the onset of the new physics interactions is so dramatic that energy, rather than accuracy, plays the crucial r\^ole in experimental searches.
As a result, electrons could be pseudo-Goldstini already at scales $\gtrsim 2 \TeV$, while light quarks require scales $\gtrsim 10 \TeV$ -- from a phenomenological point of view goldstino-compositeness is therefore on a better footing than chiral-compositeness, which requires scales $\gtrsim 40 \TeV$.
\vspace{2mm}

The scenarios with SM fermions as pseudo-Goldstini bring new interesting questions for future research. 
From an experimental point of view, these scenarios provide novel and alternative benchmarks to compare the performance of future colliders (such as FCCee, ILC or CLIC in $e^+e^-$ scattering), in which the importance of energy over accuracy is additionally emphasized.
Moreover, if other SM species are also composite, novel effects can be expected in neutral diboson pair production, which we discussed in section~\ref{sec:diboson}. These probes are very clean experimentally but they have always suffered from the lack of concrete BSM scenarios that would make them interesting: goldstino-compositeness provides a plethora of new effects that can be searched for in this type of processes (effects that, even if of the same EFT order as the ones studied so far, are complementary, better measurable and better motivated).

From a theoretical point of view goldstini-compositeness relies on  the existence of approximate and emergent supersymmetries. While proof-of-principle examples of this possibly exist, it would be interesting to set this on a firmer ground. In this roadmap, the incorporation of a solution to the hierarchy problem through  extended suspersymmetry would be an important additional target.

\section*{Acknowledgments}
We are happy to acknowledge important conversations with Zohar Komargodski, Alexander Monin, Alex Pomarol, Riccardo Rattazzi and Riccardo Torre. We thank Alexander Monin for collaboration in the early stages of this work and Simone Alioli for discussions and for providing us help with POWHEG. B.B. thanks Marco Cirelli and the LPTHE for the kind hospitality during the completion of this work.
B.B. is supported in part by the MIUR-FIRB grant RBFR12H1MW ``A New Strong Force, the origin of masses and the LHC''.

\appendix

\section*{Dileptons at LEP}
\label{dileptons}

The set of dimension-8 operators from goldstino-compositeness of leptons relevant at LEP is
\begin{align}
&\Op_{ee} = (\partial_\nu \bar{e}_R \gamma^\mu e_R)(\bar{e}_R \gamma_\mu \partial^\nu e_R) &\quad\quad&
\Op_{\ell^e e} \supset - (\partial_\nu \bar{e}_{L} \gamma^\mu e_{L}) (\partial^\nu \bar{e}_{R} \gamma_{\mu}  e_R) + \mathrm{h.c.} 
\nonumber \\
&\Op_{e\psi} = (\partial_\nu \bar{e}_R \gamma^\mu e_R)(\bar{\psi}_R \gamma_\mu \partial^\nu \psi_R) + \mathrm{h.c.}  &\quad\quad&
\Op_{\ell^e \psi} \supset - (\partial_\nu \bar{e}_{L} \gamma^\mu e_{L}) (\partial^\nu \bar{\psi}_{R} \gamma_{\mu}  \psi_R) + \mathrm{h.c.}
\nonumber\\
&\Op_{\ell^e \ell^e} \supset (\partial_\nu \bar{e}_L \gamma^\mu e_L)(\bar{e}_L \gamma_\mu \partial^\nu e_L) &\quad\quad&
\Op_{\ell^\psi e} \supset - (\partial_\nu  \bar{\psi}_{L} \gamma^\mu \psi_{L}) (\partial^\nu \bar{e}_{R} \gamma_{\mu}  e_R) + \mathrm{h.c.}
\nonumber\\
&\Op_{\ell^e \ell^\psi} \supset (\partial_\nu \bar{e}_L \gamma^\mu e_L)(\bar{\psi}_L \gamma_\mu \partial^\nu \psi_L) +\mathrm{h.c.} \, , &\quad\quad&
\label{lgops}
\end{align}
where $\psi = \mu, \tau$ and in each particular scenario $c_{ij} = 1/2F^2$. Requiring maximal $R$-symmetry implies in particular $c_{ee} = c_{e\mu} = c_{e\tau}$, $c_{\ell^e \ell^e} = c_{\ell^e \ell^\mu}= c_{\ell^e \ell^\tau}$ and $c_{\ell^e e} = c_{\ell^e \mu} = c_{\ell^\mu e} = c_{\ell^e \tau} = c_{\ell^\tau e}$.

These operators contribute to the differential dilepton cross sections,
\begin{gather}
\frac{d\sigma}{d \cos \theta}(e^+e^-\rightarrow e^+e^-)_{BSM} =  \!  \, \frac{u}{8 \pi s} \left[ \left(\frac{t^2}{s} + \frac{s^2}{t} \right) e^2 + \left(\frac{t^2}{s-m_Z^2} + \frac{s^2}{t-m_Z^2} \right) g_{e_R}^Z g_{e_L}^Z \right] c_{\ell^e e} \nonumber\\
\frac{u^3}{16 \pi s} \left[ \left(\frac{1}{s} + \frac{1}{t} \right) e^2 (c_{ee} + c_{\ell^e \ell^e})  + \left(\frac{1}{s-m_Z^2} + \frac{1}{t-m_Z^2} \right) \left( (g_{e_R}^Z)^2 c_{ee} + (g_{e_L}^Z)^2 c_{\ell^e \ell^e} \right) \right] \nonumber\\
\! + \, \frac{u^2}{32 \pi s} \left[ u^2 (c_{ee}^2 + c_{\ell^e \ell^e}^2) + 2 (t^2 + s^2 ) c_{\ell^e e}^2 \right] \, ,
\label{bhabha}\\
\frac{d\sigma}{d \cos \theta}(e^+e^-\rightarrow \psi^+\psi^-)_{BSM} =
- \frac{u^2 t}{16 \pi s} \left[ \frac{1}{s} e^2 (c_{e \psi} + c_{\ell^e \ell^\psi}) 
+ \frac{1}{s-m_Z^2} \left( (g_{e_R}^Z)^2 c_{e \psi} + (g_{e_L}^Z)^2 c_{\ell^e \ell^\psi} \right) \right] \nonumber\\
  \! + \, \frac{u^2 t}{16 \pi s} \left[ \frac{1}{s} e^2 + \frac{1}{s-m_Z^2} g_{e_R}^Z g_{e_L}^Z \right] (c_{\ell^e \psi} + c_{\ell^\psi e}) 
 \! + \, \frac{u^2 t^2}{32 \pi s} ( c_{e \psi}^2 + c_{\ell^e \ell^\psi}^2 + c_{\ell^e \psi}^2 + c_{\ell^\psi e}^2 ) \, , \quad \psi = \mu, \, \tau
\nonumber\label{dimuon}
\end{gather}
where $s$, $t$, $u$ are the Mandelstam variables ($s + t + u = 0$ in the massless approximation) and
$t = -s (1-\cos \theta)/2$ and $u = -s (1+\cos \theta)/2$, with $\theta$ is the CM scattering angle.


\end{document}